\documentclass[12pt,fleqn]{article}
\usepackage{latexsym,amssymb,graphics}
\usepackage[english]{babel}
\usepackage{epsfig,feynmp}

\makeatletter
\newif\if@preliminary
\@preliminaryfalse
\def\preliminary{\@preliminarytrue}
\def\preprintno#1{\def\@preprintno{#1}}
\def\address#1{\def\@address{\textit{#1}}}

\def\abstract#1{\def\@abstract{#1}}
\newlength\preprintnoskip
\newlength\abstractwidth
\@titlepagetrue
\renewcommand\maketitle{\begin{titlepage}%
  \let\footnotesize\small
  \hfill\parbox{\preprintnoskip}{%
  \begin{flushright}\@preprintno\end{flushright}}\hspace*{1cm}
  \vskip 60\p@
  \begin{center}%
    {\Large\bf \@title \par}\vskip 1cm%
    {\sc\@author \par}\vskip 3mm%
    {\@address \par}%
    \if@preliminary
      \vskip 2cm {\large\sf PRELIMINARY DRAFT \par \@date}%
    \fi
  \end{center}\par
  \@thanks
  \vfill
  \begin{center}%
    \parbox{\abstractwidth}{\centerline{\bf\abstractname}%
    \vskip 3mm%
    \@abstract}
  \end{center}
  \end{titlepage}%
  \setcounter{footnote}{0}%
  \let\thanks\relax\let\maketitle\relax
  \gdef\@thanks{}\gdef\@author{}\gdef\@address{}%
  \gdef\@title{}\gdef\@abstract{}\gdef\@preprintno{}
}%
\def\@citex[#1]#2{\if@filesw\immediate\write\@auxout{\string\citation{#2}}\fi
  \def\@citea{}\@cite{\@for\@citeb:=#2\do
    {\@citea\def\@citea{,\penalty\@m}\@ifundefined
       {b@\@citeb}{{\bf ?}\@warning
       {Citation `\@citeb' on page \thepage \space undefined}}%
\hbox{\csname b@\@citeb\endcsname}}}{#1}}
\def\citerange{\@ifnextchar [{\@tempswatrue\@citexr}{\@tempswafalse\@citexr[]}}
\def\@citexr[#1]#2{\if@filesw\immediate\write\@auxout{\string\citation{#2}}\fi
  \def\@citea{}\@cite{\@for\@citeb:=#2\do
    {\@citea\def\@citea{--\penalty\@m}\@ifundefined
       {b@\@citeb}{{\bf ?}\@warning
       {Citation `\@citeb' on page \thepage \space undefined}}%
\hbox{\csname b@\@citeb\endcsname}}}{#1}}
\long\def\@makecaption#1#2{%
  \vskip\abovecaptionskip
  \sbox\@tempboxa{#1: \emph{#2}}%
  \ifdim \wd\@tempboxa >\hsize
    #1: \emph{#2}\par
  \else
    \hbox to\hsize{\hfil\box\@tempboxa\hfil}%
  \fi
  \vskip\belowcaptionskip}
\newcommand{\beq}{\begin{eqnarray}}
\newcommand{\eeq}{\end{eqnarray}}
\newcommand{\nn}{\noindent}
\newcommand{\non}{\nonumber}
\newcommand{\str}{\vphantom{\bigg(}}
\newcommand{\pskip}{\vspace{\baselineskip}}
\newcommand{\s}{\\ \vspace*{-3.5mm} } 
\newcommand{\ee}{$e^+e^-$}

\newcommand{\lra}{\longrightarrow}

\makeatletter
\def\fmslash{\@ifnextchar[{\fmsl@sh}{\fmsl@sh[0mu]}}
\def\fmsl@sh[#1]#2{%
  \mathchoice
    {\@fmsl@sh\displaystyle{#1}{#2}}%
    {\@fmsl@sh\textstyle{#1}{#2}}%
    {\@fmsl@sh\scriptstyle{#1}{#2}}%
    {\@fmsl@sh\scriptscriptstyle{#1}{#2}}}
\def\@fmsl@sh#1#2#3{\m@th\ooalign{$\hfil#1\mkern#2/\hfil$\crcr$#1#3$}}
\makeatother
\def\fmfL(#1,#2,#3)#4{\put(#1,#2){\makebox(0,0)[#3]{#4}}}

\begin{document}
\baselineskip16pt   % stretch linespacing in main text

% Version with separate titlepage
\preprintno{%
hep-ph/0001169\\
DESY 99/171\\
PM/99-55\\
TTP99-48}

\title{%
  The Reconstruction of Trilinear Higgs Couplings\footnote{Proceedings, {\it Physics with a High-Luminosity} 
  $e^+ e^-\!\!$ {\it Linear Collider}, DESY/ECFA LC Workshop, DESY 99-123F.}
}
\author{%
 A.~Djouadi$^1$, W.~Kilian$^2$, M.~Muhlleitner$^3$ and 
 P.M.~Zerwas$^3$ 
}
\address{%
  $^1$Lab. de Physique Math\'{e}matique, 
        Universit\'{e} Montpellier, F-34095 Montpellier Cedex 5\\
  $^2$Institut f\"ur Theoretische Teilchenphysik, 
        Universit\"at Karlsruhe, D-76128 Karlsruhe\\
  $^3$Deutsches Elektronensynchrotron DESY, D-22603 Hamburg
}
\abstract{%
  To establish the Higgs mechanism {\it sui generis} experimentally,
  the Higgs self-interaction potential must be recon\-struc\-ted. This
  task requires the measurement of the trilinear and quadrilinear
  self-couplings, as predicted in the Standard Model or in
  supersymmetric theories. The couplings can be probed in multiple
  Higgs production at high-luminosity \ee ~linear colliders. We
  present the theoretical analysis for the production of neutral
  Higgs-boson pairs in the relevant channels of double Higgs-strahlung
  and associated multiple Higgs production.}
\maketitle
%%%%%%%%%%%%%%%%%%%%%%%%%%%%%%%%%%%%%%%%%%%%%%%%%%%%%%%%%%%%%%%%%%%%%%%%
\subsection*{1. Introduction}

{\bf 1.}  The Higgs mechanism \cite{higgs} is a cornerstone in the
electroweak sector of the Standard Model (SM) \cite{gunion}. The
electroweak gauge bosons and the fundamental matter particles acquire
masses through the interaction with a scalar field. The
self-interaction of the scalar field leads to a non-zero field
strength $v=(\sqrt{2} G_F)^{-1/2} \approx 246$~GeV in the ground state,
inducing the spontaneous breaking of the electroweak ${\rm
  SU(2)_L\times U(1)_Y}$ symmetry down to the electromagnetic ${\rm
  U(1)_{EM}}$ symmetry.\s

To establish the Higgs mechanism {\it sui generis} experimentally, the
characteristic self-energy potential of the Standard Model,
\beq 
V = \lambda \left[|\varphi|^2 -\textstyle{\frac{1}{2}} v^2 \right]^2 
\eeq 
with a minimum at $\langle \varphi \rangle_0 = v/\sqrt{2}$, must be
reconstructed once the Higgs particle will be discovered. This
experimental task requires the measurement of the trilinear and
quadrilinear self-couplings of the Higgs boson. The self-couplings are
uniquely determined in the Standard Model by the mass of the Higgs
boson which is related to the quadrilinear coupling $\lambda$ by $M_H
= \sqrt{2\lambda} v$. Introducing the physical Higgs field $H$ in the
neutral component of the doublet, $\varphi_0 = (v+H)/\sqrt{2}$, the
multiple Higgs couplings can be derived from the potential $V$:
\beq
V = \frac{M_H^2}{2}H^2+\frac{M_H^2}{2v}H^3+\frac{M_H^2}{8v^2}H^4
\eeq
The trilinear and quadrilinear vertices of the Higgs field $H$ are
given by the coefficients:
\beq
\lambda_{HHH} &=& 3 M_H^2/M_Z^2 
\qquad [{\rm unit:}\;\lambda_0=M_Z^2/v\approx33.8GeV] \\
\lambda_{HHHH} &=& 3 M_H^2/M_Z^4
\qquad [{\rm unit:}\;\lambda_0^2\,]
\eeq
For a
Higgs mass $M_H=110$~GeV, the trilinear coupling amounts to
$\lambda_{HHH} \lambda_0/ M_Z = 1.6$ for a typical energy scale $M_Z$,
whereas the quadrilinear coupling $\lambda_{HHHH} \lambda_0^2 = 0.6$
is suppressed with respect to the trilinear coupling by a factor close
to the size of the weak gauge coupling.\s

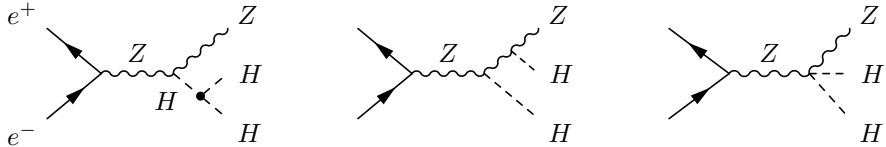
\begin{figure}
\begin{flushleft}
\underline{double Higgs-strahlung: $e^+e^-\to ZHH$}\\[1.5\baselineskip]
{\footnotesize
\unitlength1mm
\hspace{10mm}
\begin{fmfshrink}{0.7}
\begin{fmfgraph*}(24,12)
  \fmfstraight
  \fmfleftn{i}{3} \fmfrightn{o}{3}
  \fmf{fermion}{i1,v1,i3}
  \fmflabel{$e^-$}{i1} \fmflabel{$e^+$}{i3}
  \fmf{boson,lab=$Z$,lab.s=left,tens=3/2}{v1,v2}
  \fmf{boson}{v2,o3} \fmflabel{$Z$}{o3}
  \fmf{phantom}{v2,o1}
  \fmffreeze
  \fmf{dashes,lab=$H$,lab.s=right}{v2,v3} \fmf{dashes}{v3,o1}
  \fmffreeze
  \fmf{dashes}{v3,o2} 
  \fmflabel{$H$}{o2} \fmflabel{$H$}{o1}
  \fmfdot{v3}
\end{fmfgraph*}
\hspace{15mm}
\begin{fmfgraph*}(24,12)
  \fmfstraight
  \fmfleftn{i}{3} \fmfrightn{o}{3}
  \fmf{fermion}{i1,v1,i3}
  \fmf{boson,lab=$Z$,lab.s=left,tens=3/2}{v1,v2}
  \fmf{dashes}{v2,o1} \fmflabel{$H$}{o1}
  \fmf{phantom}{v2,o3}
  \fmffreeze
  \fmf{boson}{v2,v3,o3} \fmflabel{$Z$}{o3}
  \fmffreeze
  \fmf{dashes}{v3,o2} 
  \fmflabel{$H$}{o2} \fmflabel{$H$}{o1}
\end{fmfgraph*}
\hspace{15mm}
\begin{fmfgraph*}(24,12)
  \fmfstraight
  \fmfleftn{i}{3} \fmfrightn{o}{3}
  \fmf{fermion}{i1,v1,i3}
  \fmf{boson,lab=$Z$,lab.s=left,tens=3/2}{v1,v2}
  \fmf{dashes}{v2,o1} \fmflabel{$H$}{o1}
  \fmf{dashes}{v2,o2} \fmflabel{$H$}{o2}
  \fmf{boson}{v2,o3} \fmflabel{$Z$}{o3}
\end{fmfgraph*}
\end{fmfshrink}
}
\end{flushleft}
\caption{\textit{%
Subprocesses contributing to double Higgs-strahlung in the 
Standard Model at $e^+e^-$ linear colliders.
}}
\label{fig:diag}
\end{figure}

The trilinear Higgs self-coupling can be measured directly in
pair-production of Higgs particles at hadron and high-energy $e^+ e^-$
linear colliders. Several mechanisms that are sensitive to
$\lambda_{HHH}$ can be exploited for this task. Higgs pairs can be
produced through double Higgs-strahlung off $W$ or $Z$ bosons
\cite{gounaris,ilyin}, $WW$ or $ZZ$ fusion \citerange{ilyin,dicus};
moreover through gluon-gluon fusion in $pp$ collisions
\citerange{glover,dawson} and high-energy $\gamma\gamma$ fusion
\cite{ilyin,boudjema,nikia} at photon colliders.  In a precursor to
this report \cite{muhl} it was recently shown that for collider energies 
up to about 1~TeV double Higgs-strahlung is the most promising process 
for measuring the trilinear coupling:
\beq
\begin{array}{l l l c l}
\mbox{double Higgs-strahlung}& \hspace{-0.3cm} : & e^+e^- & 
\hspace{-0.3cm} \longrightarrow &\hspace{-0.1cm}  ZHH \\
\\[-0.8cm]
& & & \hspace{-0.3cm} \scriptstyle{Z} & 
\end{array} 
\eeq
As evident from Fig.~\ref{fig:diag}, the trilinear coupling is
involved only in one diagram of this process. However, the two other
diagrams are generated by the electroweak gauge interactions, and can
thus be assumed known since the Higgs-gauge coupling is measured directly 
in the basic process of single Higgs-strahlung.\s

After the decay of the Higgs bosons into $b$ and $\tau$ pairs many
reducible electroweak and QCD background processes contribute to the
final state. It has been demonstrated in careful experimental
simulations \cite{gay} and phenomenological analyses \cite{millermor} that the
signal can nevertheless be isolated in a clean form. \s
\begin{fmffile}{fd}
\\
\indent 
With values typically of the order of a few fb and below, the cross
sections are small at \ee ~linear colliders for masses of the Higgs
boson in the intermediate mass range.  High luminosities are therefore
needed to produce a sufficiently large sample of Higgs-pair events and
to isolate the signal from the background.\pskip

\nn
{\bf 2.} 
If a light Higgs boson with a mass below about 130 GeV will be
discovered, the Standard Model is likely embedded in a supersymmetric
theory. The minimal supersymmetric extension of the Standard Model
(MSSM) includes two iso-doublets of Higgs fields $\varphi_1$,
$\varphi_2$ which, after three components are absorbed to provide
longitudinal degrees of freedom to the electroweak gauge bosons, gives
rise to a quintet of physical Higgs boson states: $h$, $H$, $A$,
$H^\pm$ \cite{gunhaber}.  While an upper bound of about 130~GeV can be
derived on the mass of the light CP-even neutral Higgs boson $h$
\cite{okada,carena}, the heavy CP-even and CP-odd neutral Higgs bosons
$H$, $A$, and the charged Higgs bosons $H^\pm$ may have masses of the
order of the electroweak symmetry scale $v$ up to about 1~TeV.  This
extended Higgs system can be described by two parameters at the tree
level: one mass parameter which is generally identified with the
pseudoscalar $A$ mass $M_A$, and tan$\beta$, the ratio of the vacuum
expectation values of the two neutral fields in the two iso-doublets.
\s

The mass parameters and the couplings in the self-interaction
potential of the two Higgs doublets are affected by top and stop-loop
radiative corrections.  Radiative corrections in the one-loop leading
$M_t^4$ ap\-pro\-xi\-ma\-tion are parameterized by
\beq
\epsilon \approx \frac{3 G_F M_t^4}{\sqrt{2} \pi^2 \sin^2 \beta} 
\log \frac{\tilde{M}^2}{M_t^2} 
\eeq
where the scale of supersymmetry breaking is characterized by a common
squark-mass value $\tilde{M}$ which will be set to $1$~TeV in the numerical
analyses; if stop mixing effects are modest at the SUSY scale, they
can be accounted for by shifting $\tilde{M}^2$ in $\epsilon$ by the amount
\beq
\begin{array}{clrcl}
\tilde{M}^2 \to \tilde{M}^2 + \Delta \tilde{M}^2 &:& \Delta \tilde{M}^2 
&=& \hat{A}^2 [1-\hat{A}^2/(12\tilde{M}^2)] \\[0.1cm]
&&\hat{A} &=& A - \mu \cot\beta
\end{array}
\eeq
where $A$ and $\mu$ correspond to the trilinear coupling in the top
sector and the higgsino mass parameter in the superpotential,
respectively. The neutral CP-even Higgs boson masses, and the mixing
angle $\alpha$ in the neutral sector are given in this approximation
by

\vspace{-0.5cm}
{\footnotesize \beq
M_{h,H}^2 \!\!&=&\!\! \textstyle{\frac{1}{2}}
\left[ M_A^2+M_Z^2+\epsilon \mp
\sqrt{(M_A^2+M_Z^2+\epsilon)^2- 4M_A^2 M_Z^2 \cos^2 2\beta
   - 4\epsilon( M_A^2 \sin^2 \beta + M_Z^2 \cos^2 \beta)} \right]
\non \\
\tan 2\alpha \!\!&=&\!\! \tan 2\beta
 \frac{M_A^2 + M_Z^2}{M_A^2 - M_Z^2 +\epsilon/\cos 2\beta} \qquad
\mbox{with} \qquad  - \frac{\pi}{2} \leq \alpha \leq 0
\label{mass}
\eeq}
%\noindent 
\hspace{-0.3cm} when expressed in terms of the mass $M_A$ 
and tan $\beta$.\s

The set of trilinear couplings between the neutral physical Higgs 
bosons can be written \cite{okada,djouadi} in units of $\lambda_0$ as

\vspace{-0.5cm}
{\footnotesize \beq
\lambda_{hhh} &=& 3 \cos2\alpha \sin (\beta+\alpha) 
+ 3 \frac{\epsilon}{M_Z^2} \frac{\cos \alpha}{\sin\beta} \cos^2\alpha  
\non \\
\lambda_{Hhh} &=& 2\sin2 \alpha \sin (\beta+\alpha) -\cos 2\alpha \cos(\beta
+ \alpha) + 3 \frac{\epsilon}{M_Z^2} \frac{\sin \alpha}{\sin\beta}
\cos^2\alpha \non \\
\lambda_{HHh} &=& -2 \sin 2\alpha \cos (\beta+\alpha) - \cos 2\alpha \sin(\beta
+ \alpha) + 3 \frac{\epsilon}{M_Z^2} \frac{\cos \alpha}{\sin\beta}
\sin^2\alpha \non \\
\lambda_{HHH} &=& 3 \cos 2\alpha \cos (\beta+\alpha) 
+ 3 \frac{\epsilon}{M_Z^2} \frac{\sin \alpha}{\sin\beta} \sin^2 \alpha
\non \\
\lambda_{hAA} &=& \cos 2\beta \sin(\beta+ \alpha)+ 
\frac{\epsilon}{M_Z^2} 
\frac{\cos \alpha}{\sin\beta} \cos^2\beta \non \\
\lambda_{HAA} &=& - \cos 2\beta \cos(\beta+ \alpha) + 
\frac{\epsilon}{M_Z^2} 
\frac{\sin \alpha}{\sin\beta} \cos^2\beta
\label{coup}
\eeq}
In the decoupling limit $M_A^2 \sim M_H^2 \sim M^2_{H^\pm} \gg v^2/2$,
the self-coupling of the light CP-even neutral Higgs boson $h$
approaches the SM value.\s

\begin{figure}
\begin{center}
\epsfig{figure=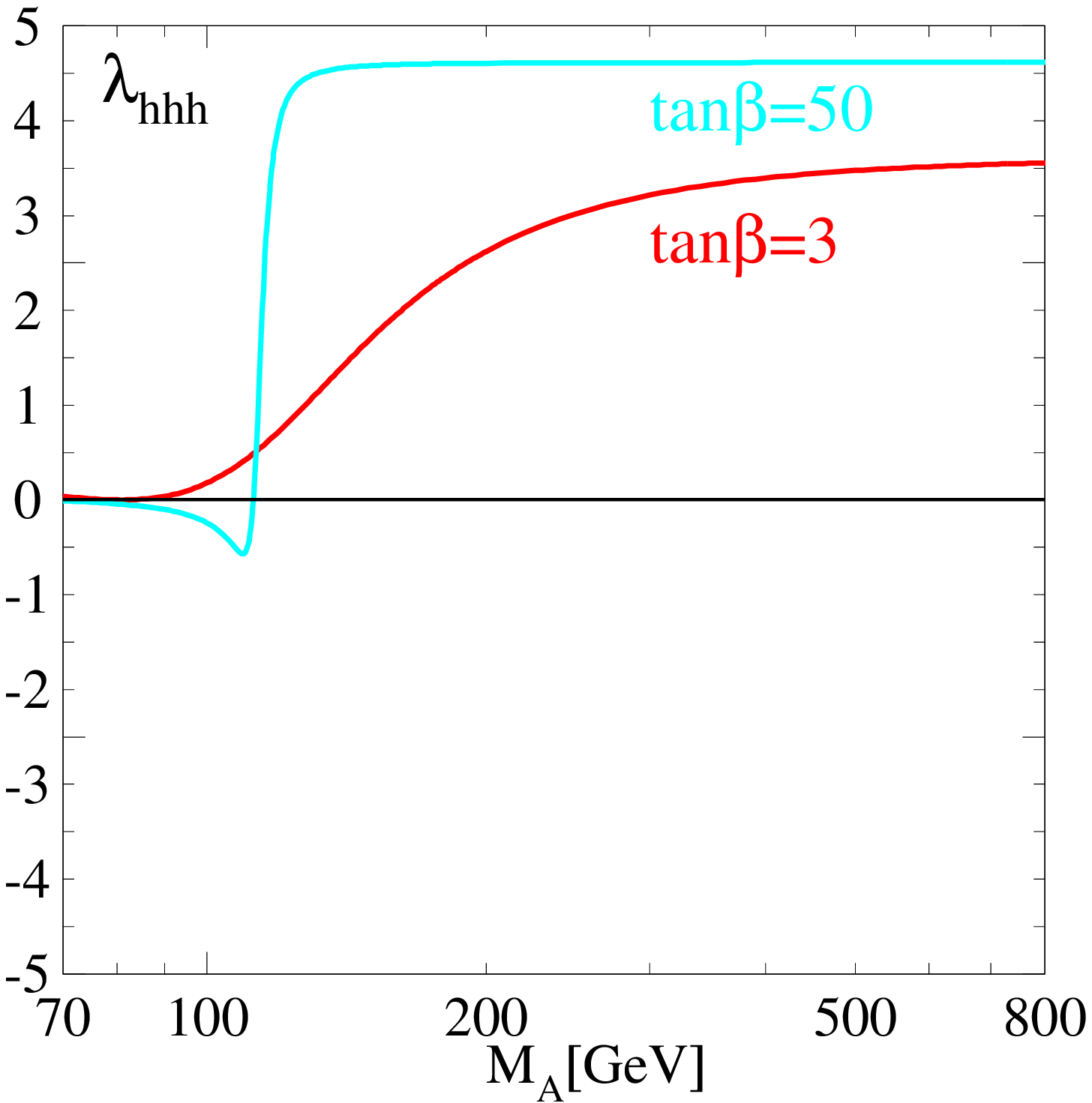,width=5.2cm}
\epsfig{figure=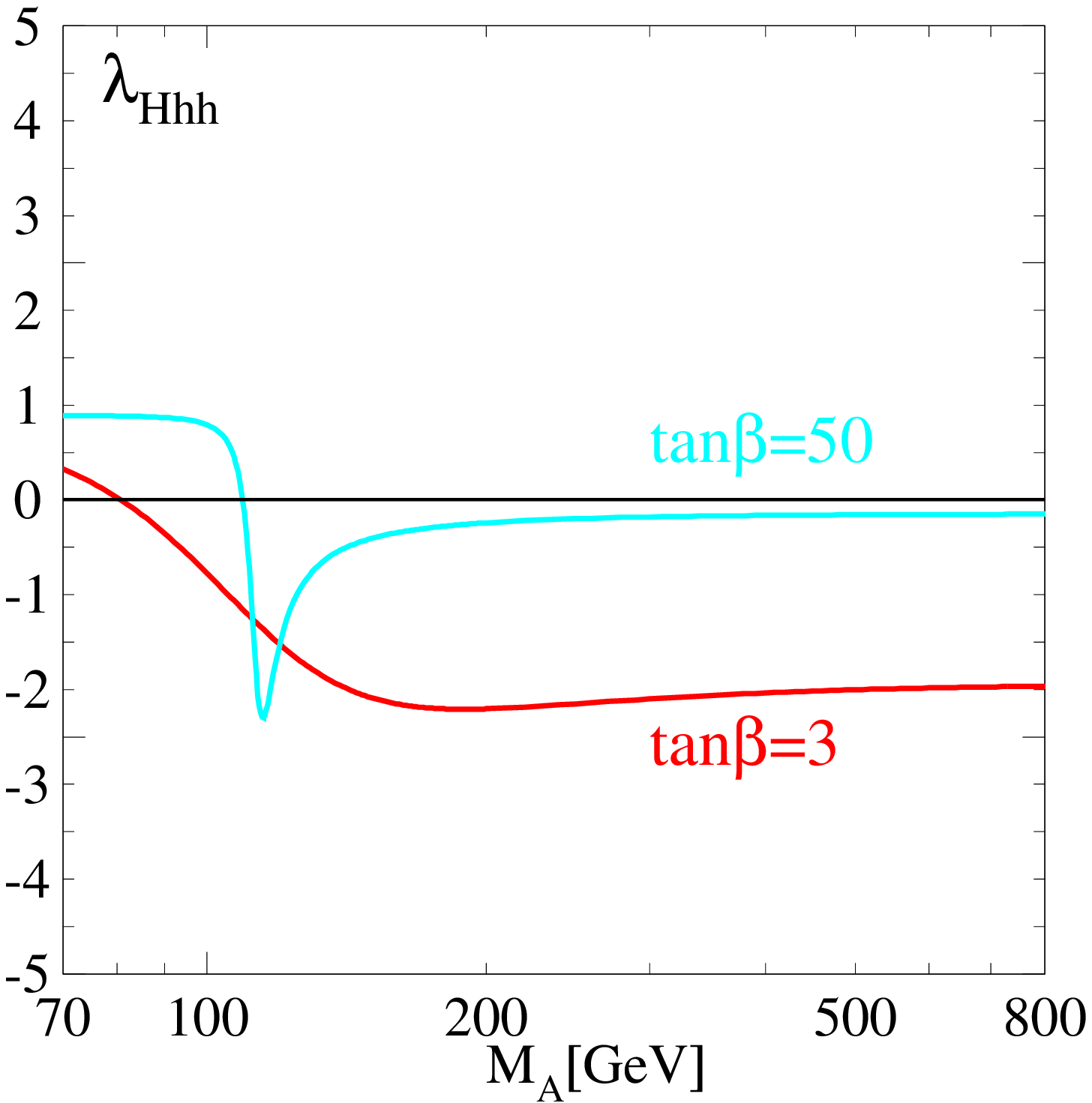,width=5.2cm}
\hspace{4mm}
\epsfig{figure=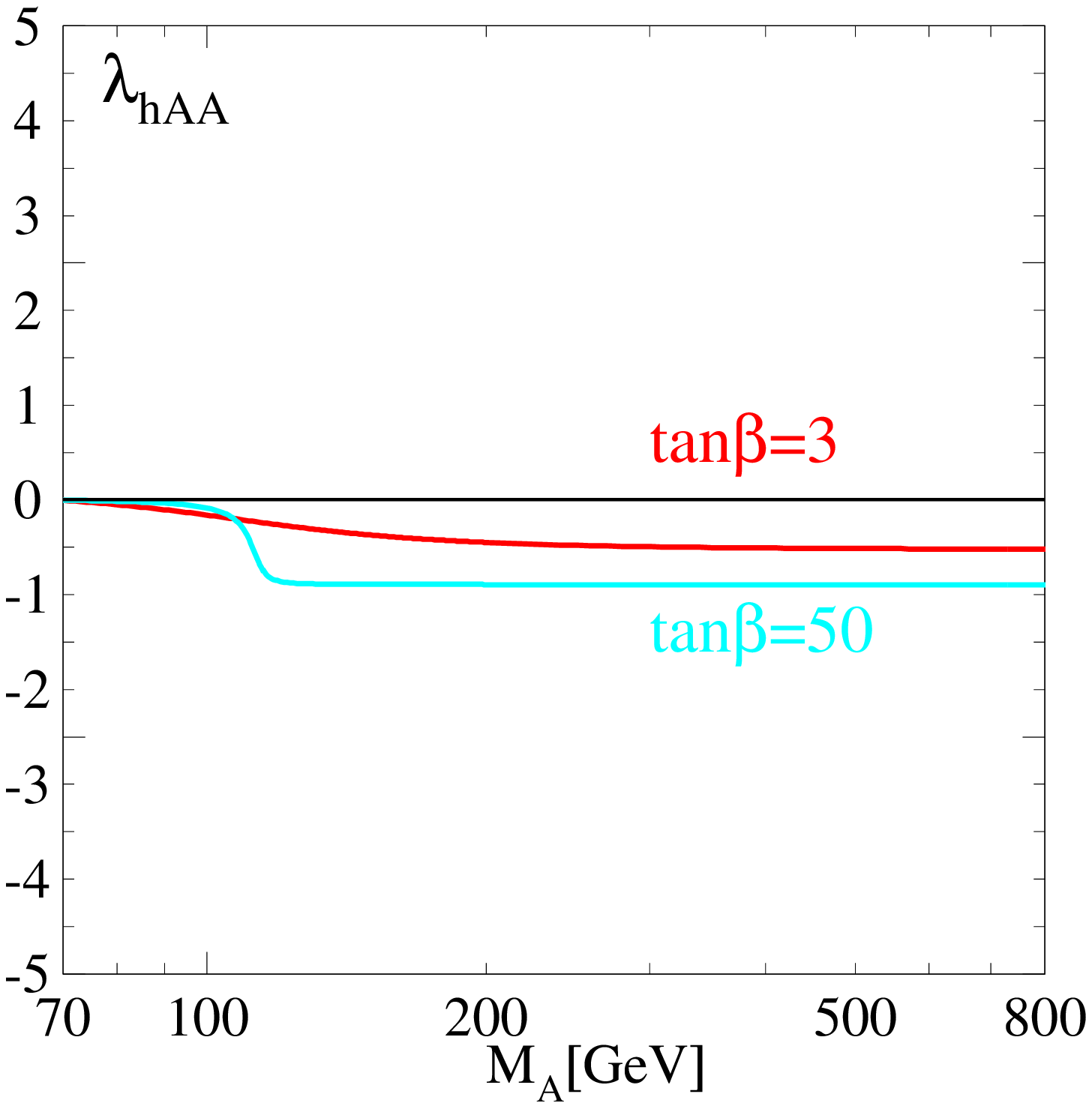,width=5.2cm}

\vspace*{5mm}

\epsfig{figure=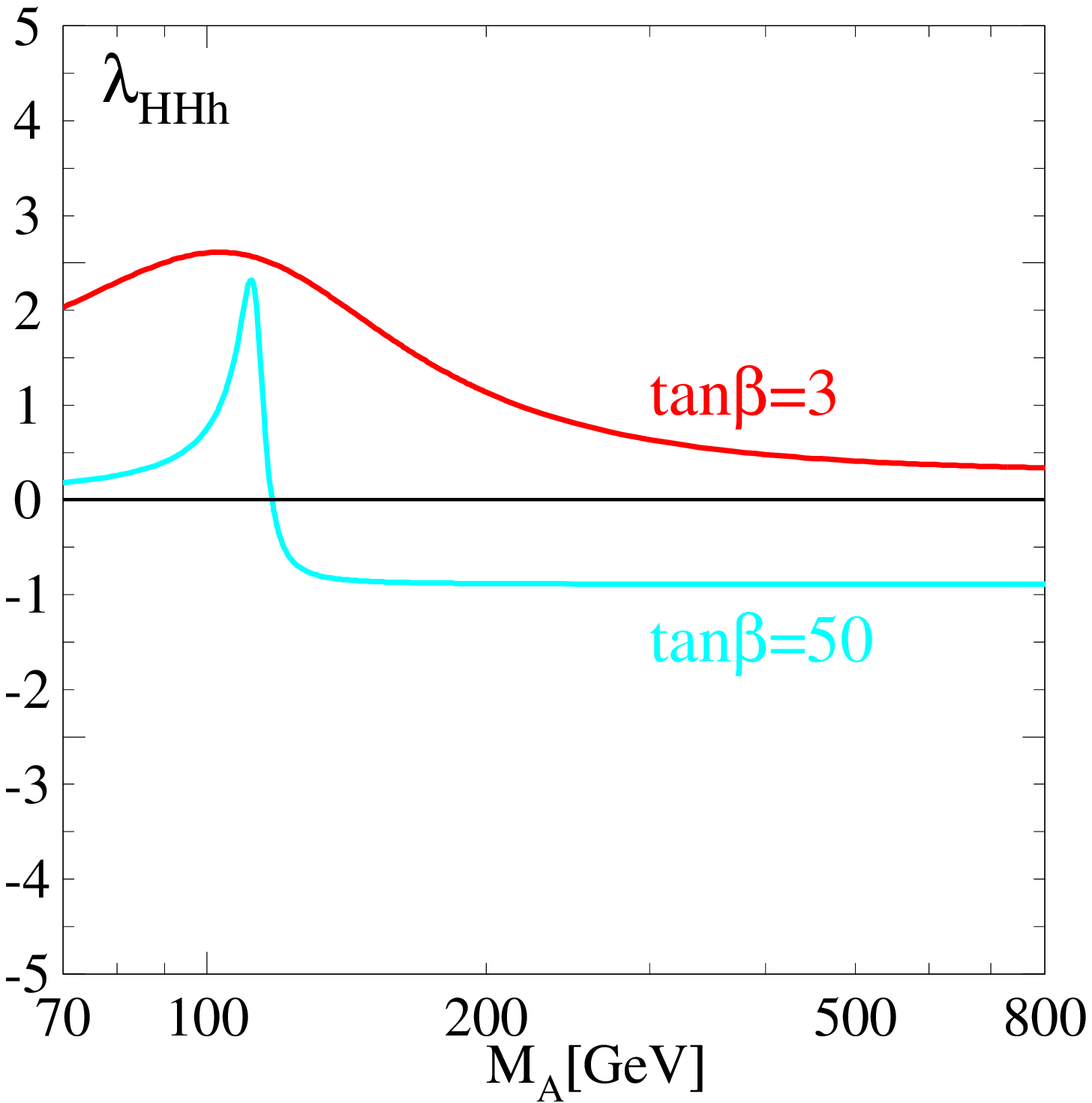,width=5.2cm}
\epsfig{figure=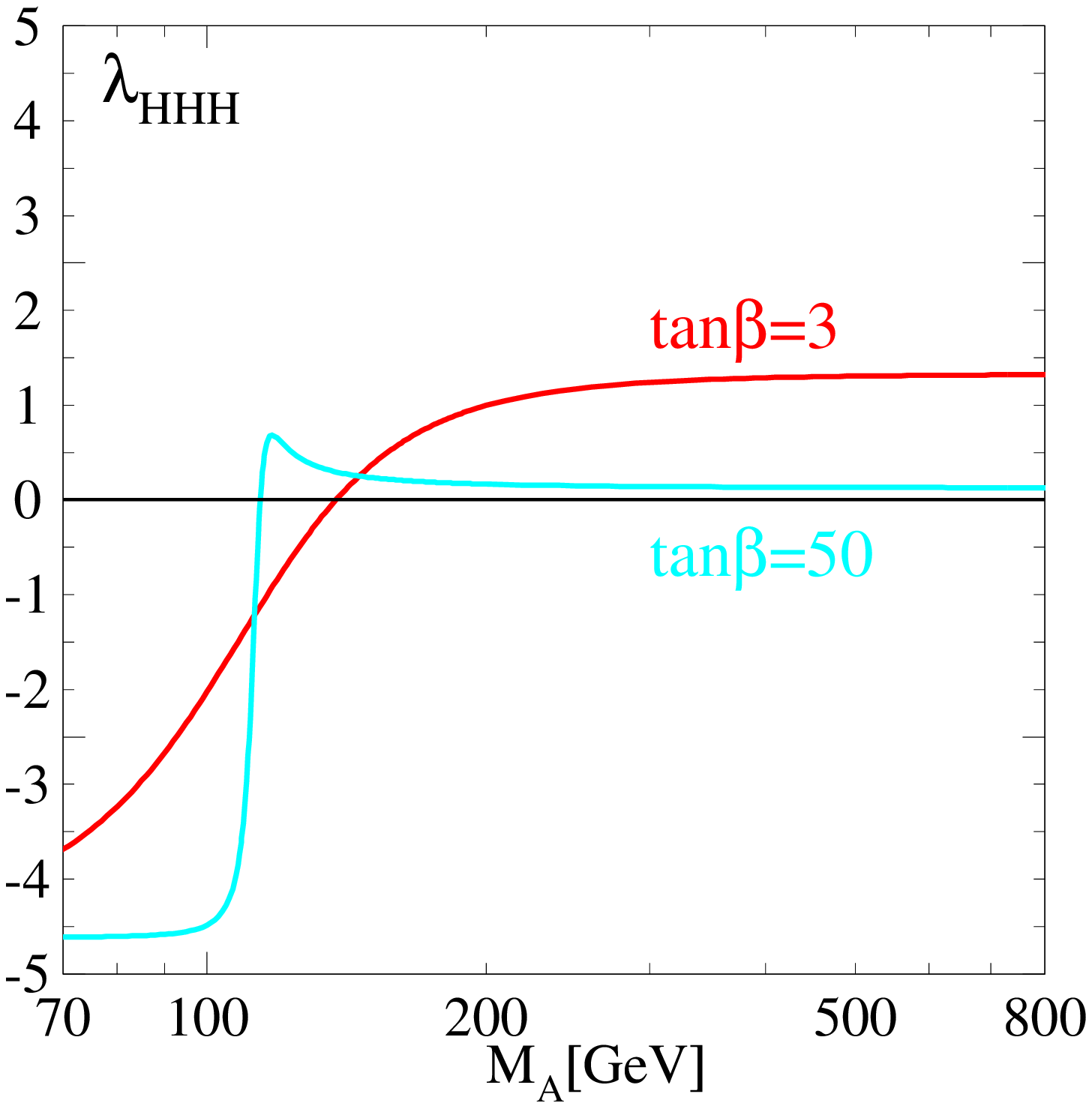,width=5.2cm}
\hspace{4mm} 
\epsfig{figure=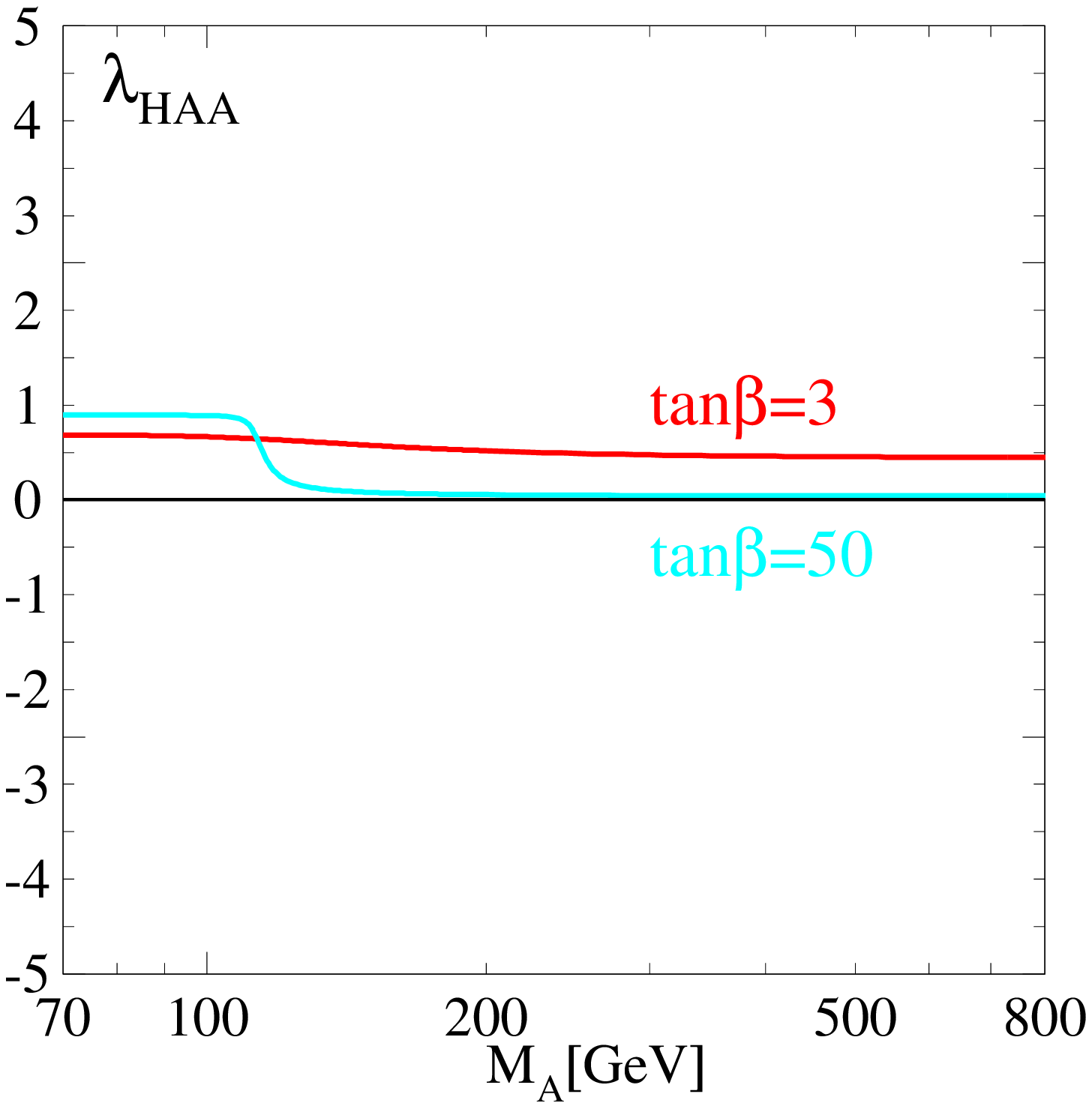,width=5.2cm}
\end{center}
Figure 2a: {\it Variation of the trilinear couplings with $M_A$ for tan$\beta 
= 3$ and $50$ in the MSSM; the region of rapid variations corresponds to the 
$h/H$ cross-over region in the CP-even sector.}
\label{fig:lambda1}
\end{figure}

\begin{figure}
\begin{center}
\epsfig{figure=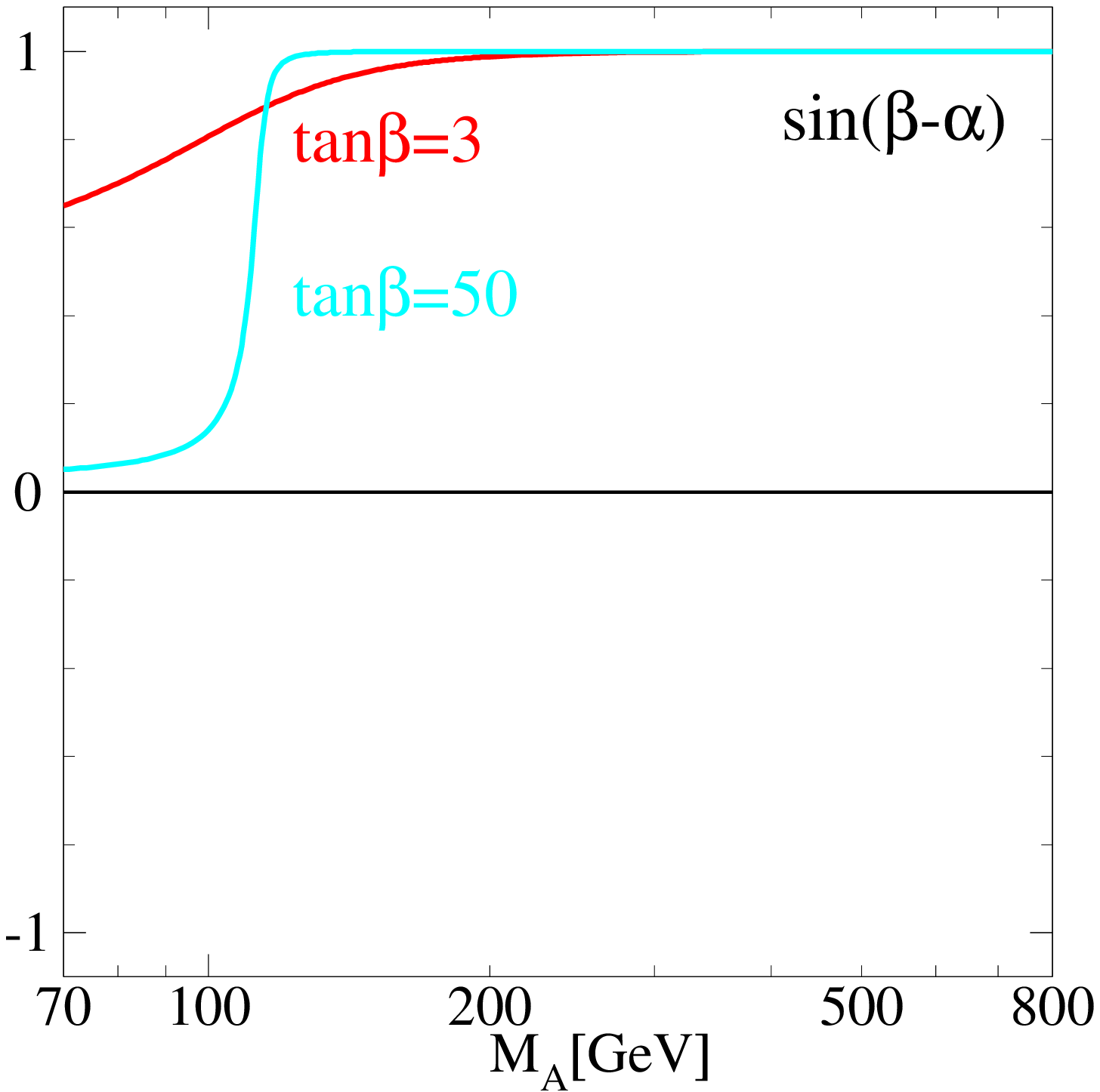,width=6cm}
\hspace{1cm}
\epsfig{figure=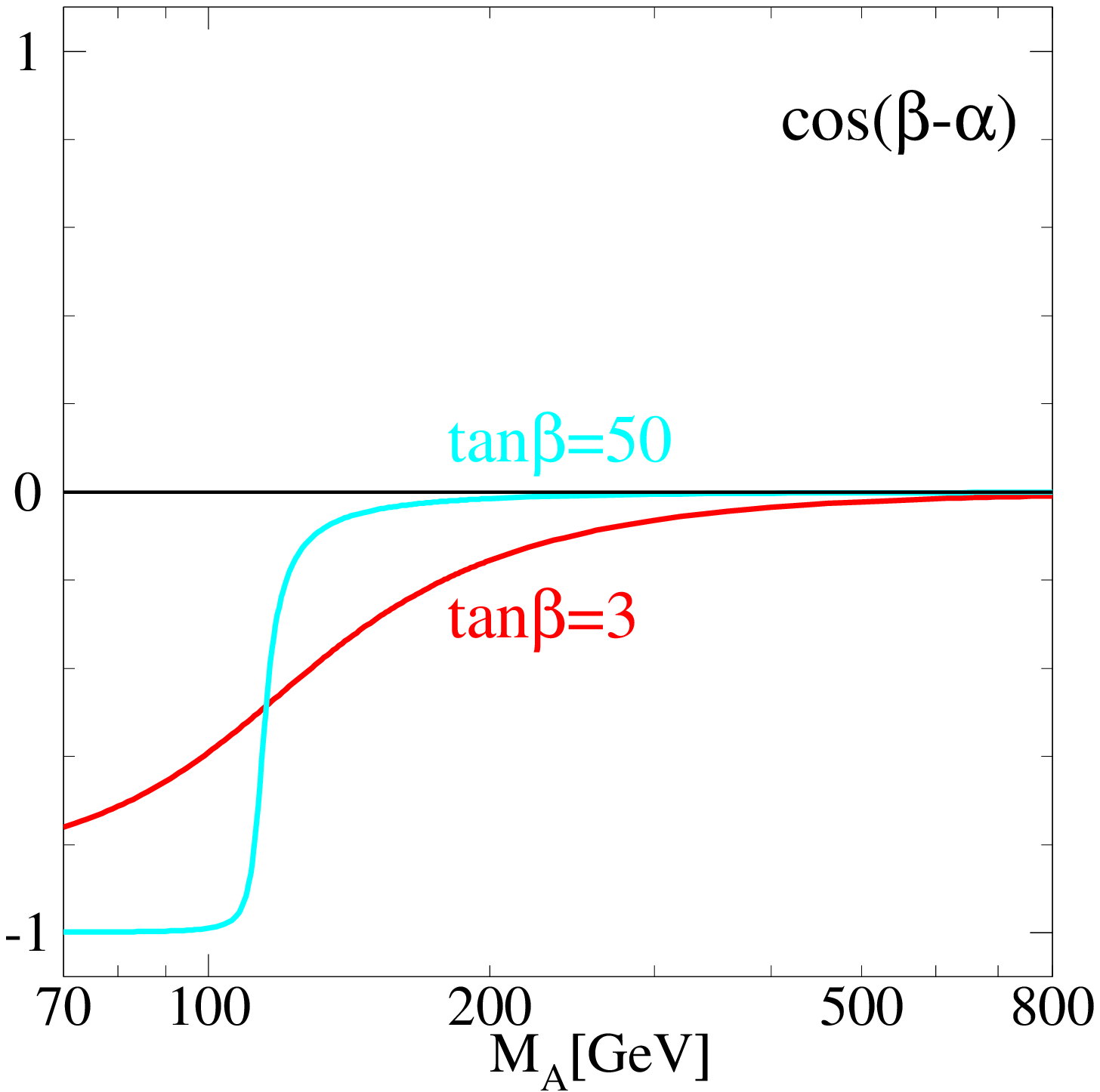,width=6cm}
\end{center}
Figure 2b: {\it ZZh and ZZH gauge couplings in units of the SM coupling.}
\label{fig:lambda2}
\end{figure}
\setcounter{figure}{2}
%
%\begin{center}
%\epsfig{figure=x111.eps,width=7.5cm}
%\hspace{1cm}
%\epsfig{figure=x112.eps,width=7.5cm}\\[2cm]
%\end{center}
%\begin{center}
%\epsfig{figure=x122.eps,width=7.5cm}
%\hspace{1cm}
%\epsfig{figure=x222.eps,width=7.5cm}
%\\[1cm]
%\end{center}
%Figure 2a: {\it Variation of the trilinear couplings between CP-even Higgs 
%bosons with $M_A$ for tan$\beta = 3$ and $50$ in the MSSM; the region 
%of rapid variations corresponds to the $h/H$ cross-over region in the 
%neutral CP-even sector.}
%\label{fig:lambda1}
%\end{figure}

%\begin{figure}
%\begin{center}
%\epsfig{figure=x133.eps,width=7.5cm}
%\hspace{1cm}
%\epsfig{figure=x233.eps,width=7.5cm}\\[2cm]
%\end{center}
%\begin{center}
%\epsfig{figure=sbma.eps,width=7.5cm}
%\hspace{1cm}
%\epsfig{figure=cbma.eps,width=7.5cm}
%\\[1cm]
%\end{center}
%Figure 2b: {\it Upper set: Variation of the trilinear scalar couplings 
%between CP-even and CP-odd Higgs bosons with $M_A$ for tan$\beta = 3$ 
%and $50$ in the MSSM. Lower set: ZZh and ZZH gauge couplings in units 
%of the SM coupling.}
%\label{fig:lambda2}
%\end{figure}
%\setcounter{figure}{2}

In the subsequent numerical analysis the complete one-loop and the
leading two-loop cor\-rec\-tions to the MSSM Higgs masses and to the
trilinear couplings are included, as presented in
Ref.~\cite{carena,hdecay}.  Mixing effects due to non-vanishing $A$,
$\mu$ parameters primarily affect the light Higgs mass; the upper
limit on $M_h$ depends strongly on the size of the mixing parameters,
raising this value for tan $\beta \gtrsim 2.5$ beyond the reach of
LEP2, cf. Ref.~\cite{carzer}. The couplings however are affected less
by higher-order corrections when evaluated for the physical Higgs
masses. The variation of the trilinear couplings with $M_A$ is shown
for two values $\tan\beta = 3$ and $50$ in Figs.~2a and
2b. The region in which the couplings vary rapidly,
corresponds to the $h/H$ cross-over region of the two mass branches in
the neutral CP-even sector, cf.~eq.~(\ref{mass}). The trilinear
couplings between $h$, $H$ and the pseudoscalar pair $AA$ are in
general significantly smaller than the trilinear couplings among the
CP-even Higgs bosons. \s
\begin{table}
\begin{center}$
\renewcommand{\arraystretch}{1.3}
\begin{array}{|l||cccc|c||ccc|}\hline
\phantom{\lambda} & 
\multicolumn{4}{|c|}{\mathrm{double\;Higgs\!-\!strahlung}} &
\multicolumn{4}{|c|}{\phantom{d} \mathrm{triple\;Higgs\!-\!production \phantom{d}}} \str \\
\phantom{\lambda i}\lambda & Zhh & ZHh & ZHH & ZAA & 
\multicolumn{2}{|c}{\phantom{d}Ahh \phantom{d}AHh} & \phantom{d}AHH & \!\! AAA \\ \hline\hline
hhh & \times & & & & \phantom{d}\times\phantom{d} & & &  \\
Hhh & \times & \times & & & \times & \times & & \\
HHh & & \times & \times & & & \times & \times & \\ 
HHH & & & \times & & & & \times & \\ 
\cline{1-6} & & & & & \multicolumn{2}{c}{\phantom{\times}} & & \\[-0.575cm]
\hline
hAA & & & & \times & \multicolumn{2}{c}{\,\,\,\times \quad\quad\, \times} & & \times \\ 
HAA & & & & \times & \multicolumn{2}{c}{\phantom{\times}\quad\quad\,\,\,\, \times}
 & \times & \times \\
\hline
\end{array}$
\end{center}
\caption{The trilinear couplings between neutral CP-even and CP-odd 
MSSM Higgs bosons which can generically be probed in double 
Higgs-strahlung and associated triple Higgs-production, are marked by 
a cross. [The matrix for WW fusion is isomorphic to the matrix for 
Higgs-strahlung.]}
\label{tab:coup}
\end{table}

In contrast to the Standard Model, resonance production of the heavy
neutral Higgs boson $H$ followed by subsequent decays $H \to hh$,
plays a dominant role in part of the parameter space for moderate
values of $\tan\beta$ and $H$ masses between 200 and 350~GeV,
Ref.~\cite{zerwas}. In this range, the branching ratio, derived from
the partial width
\beq
\Gamma [H \to hh] = 
\frac{\sqrt{2} G_F M_Z^4}{32 \pi M_H} \lambda_{Hhh}^2 \beta_h
\eeq
is neither too small nor too close to unity to be measured directly.
[The decay of either $h$ or $H$ into a pair of pseudoscalar states,
$h/H \to AA$, is kinematically not possible in the parameter range
which the present analysis is based upon; if realized, the couplings 
$\lambda_{hAA}$ and $\lambda_{HAA}$ can be determined in the same way.] 
If double Higgs production
is mediated by the resonant production of $H$, the total production
cross section of light Higgs pairs increases by about an order of
magnitude \cite{djouadi}.\s

The trilinear Higgs-boson couplings are involved in a large number of
processes at \ee ~li\-ne\-ar colliders \cite{djouadi} among which
double Higgs-strahlung and triple Higgs production are the preferred
channels \cite{muhl}:
\beq
\begin{array}{l@{:\quad}l@{\;\to\;}l l l l}
\mbox{double Higgs-strahlung} & $\ee$  & ZH_i H_j & \mathrm{and} 
& ZAA & [H_{i,j}=h,H] \\[0.2cm]
\mbox{triple Higgs production} & $\ee$ & AH_i H_j & \mathrm{and}
& AAA & 
\end{array} 
\eeq
\end{fmffile} 
\begin{figure}[t]
\begin{center}
\epsfig{figure=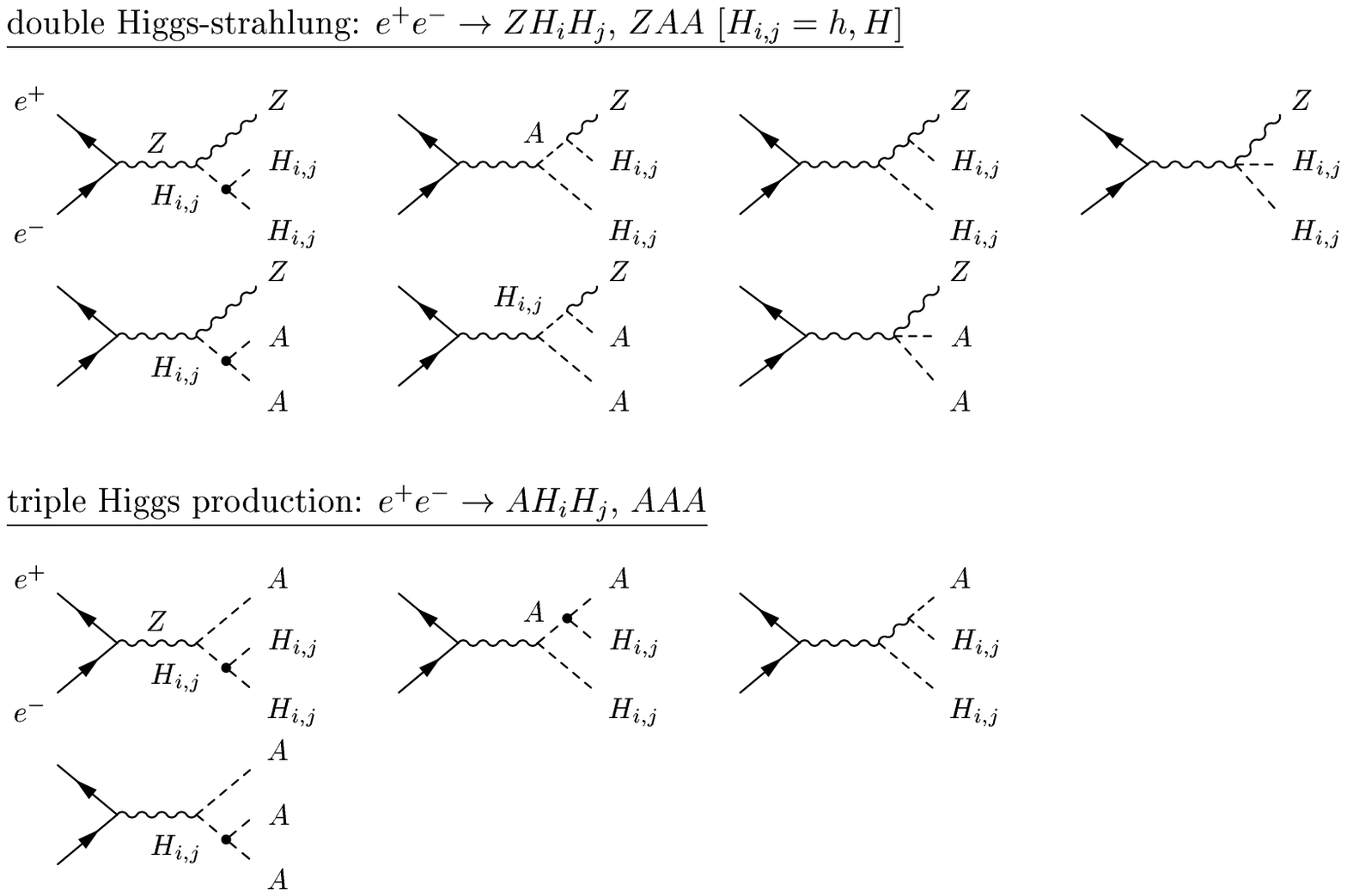,width=16cm}\\
\caption{Processes contributing to double and triple Higgs production 
involving trilinear couplings in the MSSM.}
\label{fig:graphs}
\end{center}
\end{figure}
The trilinear couplings which enter for various final states,
cf.~Fig.~\ref{fig:graphs}, are marked by a cross in the matrix
Table~\ref{tab:coup}. In the ideal case the system could be solved for
all $\lambda'$s, up to discrete ambiguities, based on double
Higgs-strahlung, $Ahh$ and triple $A$ production ["bottom-up
approach"]. This can easily be inferred from the correlation matrix
Table~\ref{tab:coup}. From $\sigma(ZAA)$ and $\sigma(AAA)$ the
couplings $\lambda(hAA)$ and $\lambda(HAA)$ can be extracted. In a
second step, $\sigma(Zhh)$ and $\sigma(Ahh)$ can be used to solve for
$\lambda(hhh)$ and $\lambda(Hhh)$; subsequently, $\sigma(ZHh)$ for
$\lambda(HHh)$; and, finally, $\sigma(ZHH)$ for $\lambda(HHH)$. The
remaining triple Higgs cross sections $\sigma(AHh)$ and $\sigma(AHH)$
could provide additional redundant information on the trilinear
couplings.\s

In practice, not all the cross sections will be large enough to be
accessible experimentally, preventing the straightforward solution for
the complete set of couplings. In this situation however the reverse
direction can be followed ["top-down approach"]. The trilinear Higgs
couplings can stringently be tested by comparing the theoretical
predictions of the cross sections with the experimental results for
the accessible channels of double and triple Higgs production. \s

If, as expected in the MSSM, the couplings $hAA$ and $HAA$ are very
small, the $ZAA$ and $AAA$ final states can be left out of the
analysis. This conclusion can be checked experimentally in a
model-independent way, assuming nothing but the knowledge of
pre-determined gauge boson-Higgs couplings. The system is then reduced
to the trilinear couplings among the CP-even Higgs bosons $h$, $H$ [in
the double-line box of Table \ref{tab:coup}] which can be measured in
the analysis chain outlined in the previous paragraph, based solely on
double Higgs-strahlung $ZH_i H_j$ and triple Higgs-production $Ahh$.
The $hH_i H_j$ couplings involving the light Higgs boson $h$ with any
combination of CP-even Higgs bosons can thus be determined in total.
\s

The processes \ee$\to ZH_i A$ [$H_i=h,$ $H$] of mixed CP-even/CP-odd
Higgs final states are generated through gauge interactions alone,
mediated by virtual $Z$ bosons decaying to the CP even--odd Higgs
pair, $Z^* \to H_i A$. These parity-mixed processes do not involve
trilinear Higgs-boson couplings.  \pskip

\nn
{\bf 3.}
In this report we summarize the results of Ref.~\cite{muhl} for the
production of Higgs boson pairs in the Standard Model and in the
minimal supersymmetric extension. Other aspects have been discussed in
Refs.~\cite{osland,dubinin}.  The comparison with LHC channels has been
presented in Refs.~\cite{muehl,lhc}. The analyses have been carried out
for $e^+ e^-$ linear colliders \cite{acco}, which are currently
designed for an initial phase in the range $\sqrt{s} = 500$~GeV to
1~TeV. The small cross sections require high luminosities as foreseen
in the TESLA design with targets of $\int {\cal L} = 300$ and
500~fb$^{-1}$ {\it per annum} for $\sqrt{s} = 500$ and $800$~GeV,
respectively \cite{brink}. \s

The report is divided into two parts. In Section 2 we discuss the
measurement of the trilinear Higgs coupling in the Standard Model for
double Higgs-strahlung at \ee ~linear colliders. In Section 3 this
program, including the triple Higgs production, is extended to the
Minimal Supersymmetric Standard Model MSSM.

%%%%%%%%%%%%%%%%%%%%%%%%%%%%%%%%%%%%%%%%%%%%%%%%%%%%%%%%%%%%%%%%%%%%%%%%
\subsection*{2. Higgs Pair--Production in the Standard Model}

\subsubsection*{2.1 Double Higgs-strahlung}

\begin{figure}
\begin{center}
\epsfig{figure=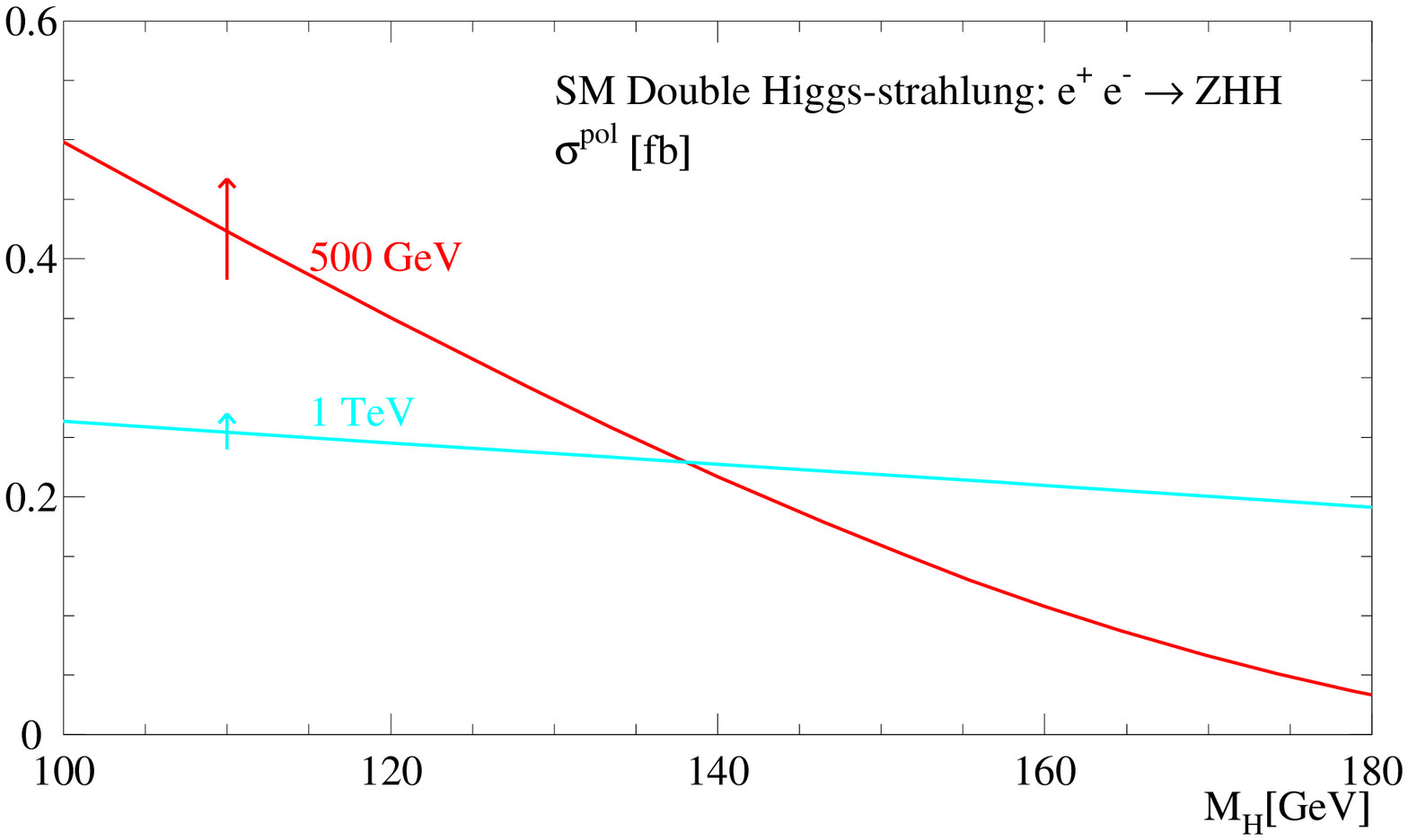,width=13cm}
\\
\end{center}
Figure 4a: {\it The cross section for double Higgs-strahlung in the SM 
at two collider energies: $500$~GeV and $1$~TeV. The 
electron/positron beams are taken oppositely polarized. The vertical 
arrows correspond to a variation of the trilinear Higgs coupling from 
$0.8$ to $1.2$ of the SM value.}
\label{fig:SM1}
\end{figure}
\begin{figure}
\begin{center}
\epsfig{figure=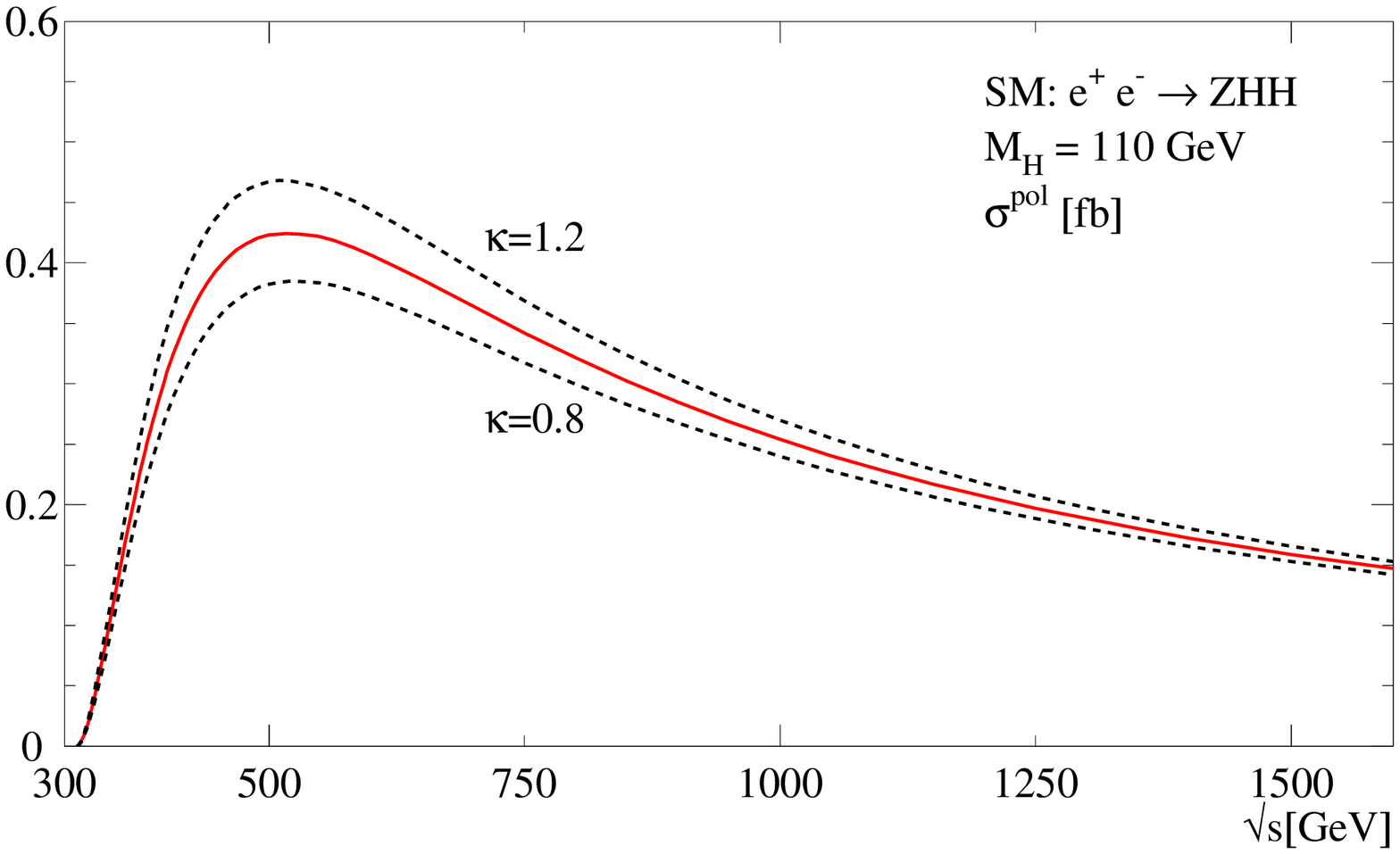,width=13cm}
\end{center}
Figure 4b: {\it The energy dependence of the cross section for double 
Higgs-strahlung for a fixed Higgs mass $M_H=110$~GeV. The full line 
corresponds to the trilinear Higgs coupling of the Standard Model. 
The variation of the cross section for modified trilinear couplings 
$\kappa\lambda_{HHH}$ is indicated by the dashed lines.}
\end{figure}
\setcounter{figure}{4}

The (unpolarized) differential cross section for the process of double
Higgs-strahlung \ee $\to ZHH$, cf.~Fig.~\ref{fig:diag}, can be cast
into the form \cite{djouadi}
\beq 
\frac{d \sigma [e^+ e^- \to ZHH]}{d x_1 d x_2} = 
\frac{\sqrt{2} G_F^3 M_Z^6}{384 \pi^3 s}
\frac{v_e^2 + a_e^2}{(1- \mu_Z)^2}\, {\cal Z}(x_1,x_2) 
\eeq 
after the angular dependence is integrated out. The vector and
axial-vector $Z$ charges of the electron are defined as usual, by $v_e
= -1 + 4\sin^2 \theta_W$ and $a_e = -1$. $x_{1,2} = 2
E_{1,2}/\sqrt{s}$ are the scaled energies of the two Higgs particles,
$x_3 = 2 - x_1 -x_2$ is the scaled energy of the $Z$ boson, and $y_i =
1 - x_i$; the square of the reduced masses is denoted by $\mu_i =
M_i^2/s$, and $\mu_{ij}=\mu_i-\mu_j$. In terms of these variables,
the coefficient ${\cal Z}$ may be written as:
\beq 
{\cal Z} =  {\mathfrak z}^2 f_0 +
\frac{1}{4 \mu_Z (y_1+\mu_{HZ})} \left[ 
\frac{f_1}{y_1+\mu_{HZ}} + \frac{f_2}{y_2+\mu_{HZ}} 
+ 2\mu_Z {\mathfrak z} f_3 \right] 
+ \Bigg\{ y_1 \leftrightarrow y_2 \Bigg\} 
\eeq
with
\beq
{\mathfrak z} = \frac{\lambda_{HHH}}{y_3-\mu_{HZ}}
 + \frac{2}{y_1+\mu_{HZ}} + 
\frac{2}{y_2+\mu_{HZ}} + \frac{1}{\mu_Z}
\eeq
The coefficients $f_i$ are listed in Ref.~\cite{muhl}. The first term
in the coefficient ${\mathfrak z}$ includes the trilinear coupling
$\lambda_{HHH}$. The other terms are related to sequential
Higgs-strahlung amplitudes and the 4-gauge-Higgs boson coupling; the
individual terms can easily be identified by examining the
characteristic propagators. \s

Since double Higgs-strahlung is mediated by s-channel $Z$-boson
exchange, the cross section doubles if oppositely polarized electron
and positron beams are used. \s

The cross sections for double Higgs-strahlung in the intermediate mass
range are presented in Fig.~4a for total $e^+ e^-$ energies of
$\sqrt{s} = 500$~GeV and 1~TeV. The cross sections are shown for
polarized electrons and positrons, $\lambda_{e^-} \lambda_{e^+} = -1$.
As a result of the scaling behavior, the cross section for double
Higgs-strahlung decreases with rising energy beyond the threshold
region. The cross section increases with rising trilinear
self-coupling in the vicinity of the SM value. The sensitivity to the
$HHH$ self-coupling is demonstrated in Fig.~4b by varying the
trilinear coupling $\kappa\lambda_{HHH}$ within the range $\kappa =
0.8$ and $1.2$; the sensitivity is also illustrated by the vertical
arrows in Fig.~4a for a variation of $\kappa$ in the same range.
Evidently the cross section $\sigma($\ee$\to ZHH)$ is sensitive to the
value of the trilinear coupling; the sensitivity is not swamped by the
irreducible background diagrams involving only the Higgs-gauge
couplings. While the irreducible background diagrams become more
important for rising energies, the sensitivity to the trilinear Higgs
coupling is very large just above the kinematical threshold for the
$ZHH$ final state. Near the threshold the value of the propagator of the
intermediate virtual Higgs boson connecting to the two real Higgs
bosons through $\lambda_{HHH}$ in the final state, is maximal. The
maximum cross section for double Higgs-strahlung is reached at
energies $\sqrt{s}\sim 2M_H+M_Z+200$~GeV, i.e. for Higgs masses in the
lower part of the intermediate range at $\sqrt{s} \sim 500$~GeV.\s

Below 1~TeV one can always find collider energies at which the cross
section for Higgs-strahlung $e^+e^- \to ZHH$ is larger than the cross
section for $WW$ fusion of two Higgs bosons, $e^+e^- \to \bar{\nu}_e
\nu_e HH$. However, for collider energies above 1~TeV the logarithmic
increase of the $t$-channel fusion process dominates over the
Higgs-strahlung mechanism which scales in the energy.\s

In recent experimental simulations of the Higgs-strahlung process it
has been shown that the signal for two-Higgs boson production can be
extracted \cite{gay} despite the multi-channel reducible background
\cite{millermor}. A sensitivity better than $20 \%$ can be expected
for the measurement of the trilinear Higgs self-coupling in the lower
part of the intermediate Higgs mass range of the Standard Model.\s

The complete reconstruction of the Higgs potential in the Standard
Model requires the measurement of the quadrilinear coupling
$\lambda_{HHHH}$, too. This coupling is sup\-pres\-sed relative to the
trilinear coupling by a factor which is effectively of the order of
the weak gauge coupling for masses in the lower part of the
intermediate Higgs mass range. Access to the quadrilinear coupling can
be obtained directly only through the production of three Higgs
bosons: \ee $\lra ZHHH$. However, this cross section is strongly
reduced by three orders of magnitude compared to the corresponding
double-Higgs channel. As argued before, the signal amplitude involving
the four-Higgs coupling [as well as the irreducible Higgs-strahlung
amplitudes] is suppressed, leading to a reduction by a factor
$[\lambda_{HHHH}^2\lambda_0^4/16\pi^2]/[\lambda_{HHH}^2\lambda_0^2/M_Z^2]
\sim 10^{-3}$. Irreducible background diagrams are suppressed by a
ratio of similar size. Moreover, the phase space is reduced by the
additional heavy particle in the final state. A few illustrative
examples of cross sections for triple Higgs-strahlung are listed in
Table~\ref{tab:quadri}.

\begin{table}
\begin{center}$
\begin{array}{|rl|lrl|}\hline
\multicolumn{2}{|c|}{\sqrt{s}=1 \mathrm{TeV}} &
\multicolumn{3}{|c|}{\sigma($\ee$\to ZHHH)[\mathrm{ab}]} 
\str \\
\hline 
M_H\!=\!\!\!\!&110\;\mathrm{GeV} &
\hspace{0.5cm} 0.44 & [0.41/\!\!\!\!& 0.46] \str \\ 
\phantom{val} & 150\;\mathrm{GeV} &
\hspace{0.5cm}0.34 & [0.32/\!\!\!\!& 0.36] \str \\ 
\phantom{val} & 190\;\mathrm{GeV} &
\hspace{0.5cm}0.19 & [0.18/\!\!\!\!& 0.20] \str \\ \hline 
\end{array}$
\end{center}
\caption{Representative values for triple SM Higgs-strahlung 
(unpolarized beams). The sensitivity to the quadrilinear coupling is 
illustrated by the variation of the cross sections when 
$\lambda_{HHHH}$ is altered by factors $1/2$ and $3/2$, as indicated 
in the square brackets.}
\label{tab:quadri}
\end{table}

%%%%%%%%%%%%%%%%%%%%%%%%%%%%%%%%%%%%%%%%%%%%%%%%%%%%%%%%%%%%%%%%%%%%%%%%
\subsection*{3. The Supersymmetric Higgs Sector}

A large ensemble of Higgs couplings are present in supersymmetric
theories. Even in the minimal realization MSSM, six different
trilinear couplings $hhh$, $Hhh$, $HHh$, $HHH$, $hAA$, $HAA$ are
generated among the neutral particles, and many more quadrilinear
couplings. Since in major parts of the MSSM parameter
space the Higgs bosons $H$, $A$, $H^\pm$ are quite heavy, we will
focus primarily on the production of light neutral pairs $hh$, yet the
production of heavy Higgs bosons will also be discussed where
appropriate.  The channels in which trilinear Higgs couplings can be
probed in \ee ~collisions, have been catalogued in
Table~\ref{tab:coup}.\s

Barring the exceptional case of very light pseudoscalar $A$ states,
$\lambda_{Hhh}$ is the only trilinear coupling that may be measured in
resonance decays, $H\to hh$, while all the other couplings
must be accessed in continuum pair production. The relevant mechanisms
have been ca\-te\-go\-ri\-zed in Fig.~\ref{fig:graphs} for double
Higgs-strahlung and associated triple Higgs production.\s

\subsubsection*{3.1 Double Higgs-strahlung}

The (unpolarized) cross section for double Higgs-strahlung, $e^+ e^-
\to Zhh$, is modified \cite{muhl,djouadi,osland} with regard to the
Standard Model by $H$,$A$ exchange diagrams,
cf.~Fig.~\ref{fig:graphs}:
\beq
\frac{d \sigma [e^+ e^- \to Zhh]}{d x_1 d x_2} &=& 
\frac{\sqrt{2} G_F^3 M_Z^6}{384 \pi^3 s} 
\frac{v_e^2 + a_e^2}{(1- \mu_Z)^2}\, {\cal Z}_{11}(x_1,x_2)
\label{zhh1}
\eeq
\hspace{-0.1cm} with
\vspace{-0.5cm}
{\footnotesize 
\beq
{\cal Z}_{11} &=&  {\mathfrak z}^2_{11} f_0 + 
\frac{{\mathfrak z}}{2} \left[ 
\frac{\sin^2 (\beta-\alpha) f_3}{y_1 + \mu_{1Z}} + 
\frac{\cos^2 (\beta-\alpha) f_3}{y_1 + \mu_{1A}} \right]  
+ \frac{\sin^4 (\beta-\alpha)}{4\mu_Z (y_1+\mu_{1Z})} \left[ 
\frac{f_1}{y_1+\mu_{1Z}} + \frac{f_2}{y_2+\mu_{1Z}} \right] \non\\
&+& \frac{\cos^4 (\beta-\alpha)}{4\mu_Z (y_1+\mu_{1A})} \left[ 
\frac{f_1}{y_1+\mu_{1A}} + \frac{f_2}{y_2+\mu_{1A}} \right] 
+ \frac{\sin^2 2(\beta-\alpha)}{8\mu_Z (y_1+\mu_{1A})} \left[ 
\frac{f_1}{y_1+\mu_{1Z}} + \frac{f_2}{y_2+\mu_{1Z}} \right] \non\\
&+& \Big\{ y_1 \leftrightarrow y_2 \Big\}
\label{zhh2}
\eeq}
\hspace{-0.3cm} and
\vspace{-0.5cm}
{\footnotesize 
\beq
{\mathfrak z}_{11} = \left[ 
\frac{\lambda_{hhh}\sin(\beta-\alpha)}{y_3-\mu_{1Z}}
+ \frac{\lambda_{Hhh}\cos(\beta-\alpha)}{y_3 - \mu_{2Z}} \right] 
+ \frac{2 \sin^2(\beta-\alpha)}{y_1+\mu_{1Z}} 
+ \frac{2 \sin^2(\beta-\alpha)}{y_2+\mu_{1Z}} 
+ \frac{1}{\mu_Z}
\label{zhh3}
\eeq}
\hspace{-0.3cm} The notation follows the Standard Model, with
$\mu_1=M_h^2/s$ and $\mu_2=M_H^2/s$. In parameter ranges in which the
heavy neutral Higgs boson $H$ or the pseudoscalar Higgs boson $A$
becomes resonant, the decay widths are implicitly included by shifting
the masses to complex Breit-Wigner values. \s
\begin{figure}
\begin{center}
\epsfig{figure=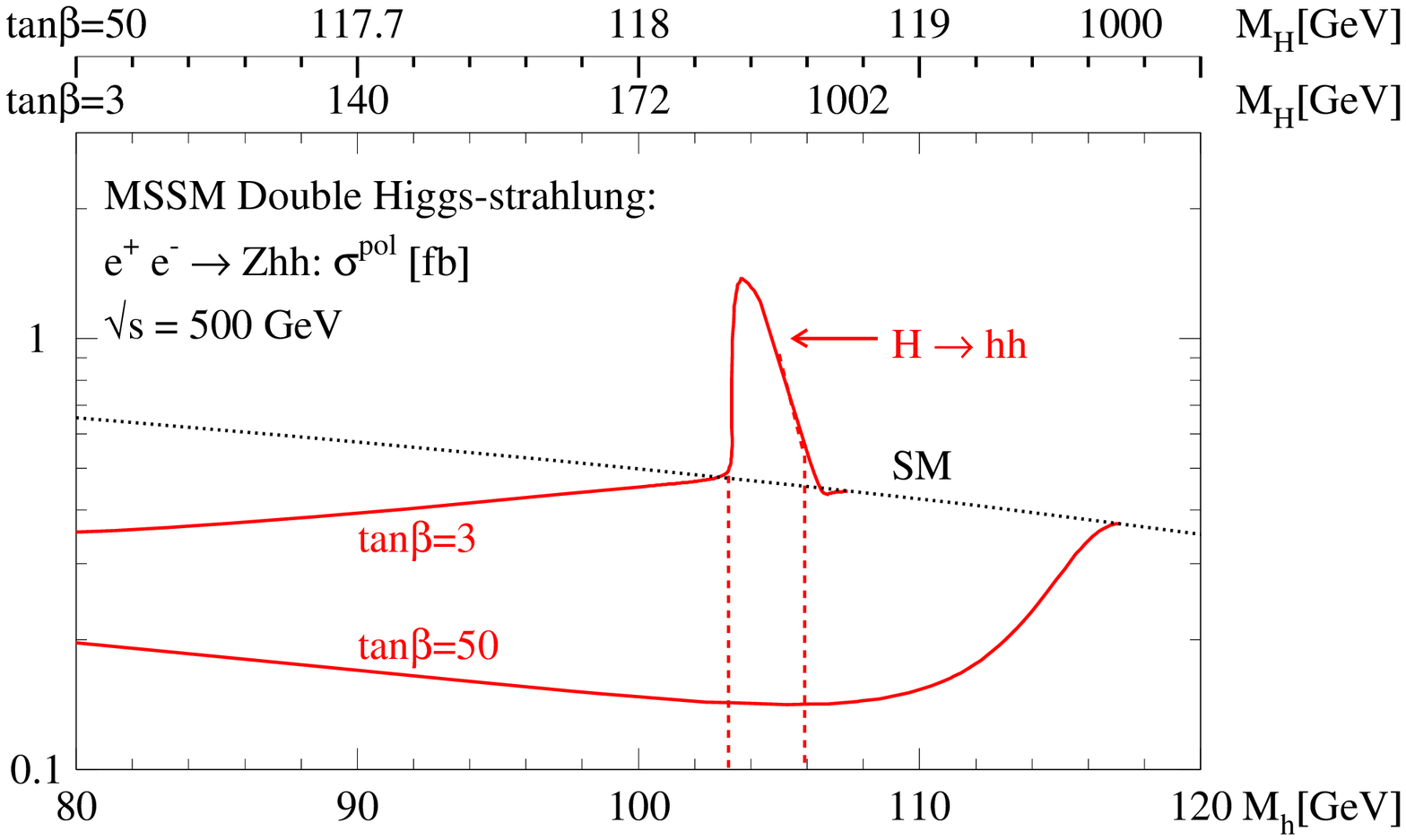,width=13cm}
\caption{
  Total cross sections for MSSM $hh$ production via double
  Higgs-strahlung at $e^+e^-$ linear colliders for $\tan\beta =3$, $50$ 
  and $\sqrt{s}=500\;\mathrm{GeV}$, including mixing effects 
  ($A = 1$~TeV, $\mu=-1/1$~TeV for $\tan\beta=3/50$). 
  The dotted line indicates the SM cross section.}
\label{fig:NLC/SUSY}
\end{center}
\end{figure}

The total cross sections are shown in Fig.~\ref{fig:NLC/SUSY} for the
$e^+ e^-$ collider energy $\sqrt{s}= 500$ GeV. The parameter $\tan
\beta$ is chosen to be 3 and 50, and the mixing parameters $A = 1$~TeV
and $\mu = -1$~TeV and $1$~TeV, respectively. If $\tan \beta$ and the
mass $M_h$ are fixed, the masses of the other heavy Higgs bosons are
predicted in the MSSM \cite{zerwas}. Since the vertices are suppressed
by $\sin/\cos$ functions of the mixing angles $\beta$ and $\alpha$,
the continuum $hh$ cross sections are suppressed compared to the
Standard Model. The size of the cross sections increases for moderate
$\tan \beta$ by nearly an order of magnitude if the $hh$ final state
can be generated in the chain $e^+ e^- \to ZH \to Zhh$ via resonant
$H$ Higgs-strahlung. If the light Higgs mass approaches the upper
limit for a given value of $\tan \beta$, the decoupling theorem drives
the cross section of the supersymmetric Higgs boson back to its
Standard Model value since the Higgs particles $A$, $H$, $H^\pm$
become asymptotically heavy in this limit. As a result of the
decoupling theorem, resonance production is not effective for large
tan$\beta$.  If the $H$ mass is large enough to allow for decays to $hh$
pairs, the $ZZH$ coupling is already too small to generate a sizable
cross section.\s

While the basic structure for the cross sections of the other $ZH_i
H_j$ [$H_{i,j}=h$, $H$] final states remains the same, the complexity
increases due to unequal masses of the final-state particles. 
The double differential cross section of the process $e^+ e^- \to ZH_i
H_j$ is given for unpolarized beams by the expression
\beq
\frac{d\sigma[e^+ e^- \to ZH_i H_j]}{dx_1 dx_2} &=&  
\frac{ \sqrt{2} \, G_F^3 \, M_Z^6}{ 384\, \pi^3 s\,} 
\frac{ v_e^2+a_e^2}{(1-\mu_Z)^2} \  {\cal Z}_{ij}(x_1,x_2) 
\label{eq:sigma}
\eeq
The coefficients ${\cal Z}_{ij}$ in the cross sections can be written as
{\footnotesize
\beq
{\cal Z}_{ij} &=& {\mathfrak z}^2_{ij}\,f_0 
+ \frac{{\mathfrak z}_{ij}}{2} \, \left[
  \frac{ d_i d_j \, f_3 }{y_1+\mu_{iZ}} 
+ \frac{ c_i c_j \, f_3}{y_1+\mu_{iA}} 
   \right] +\frac{ (d_i d_j)^2}{4\mu_Z (y_1+\mu_{iZ})}
\left[\frac{f_1}{y_1+\mu_{iZ}}+ \frac{f_2}{y_2+\mu_{jZ}} \right]\non \\
&+&  \!\!\!
\frac{ (c_i c_j )^2 }{4\mu_Z (y_1+\mu_{iA})}
\left[
\frac{f_1}{y_1+\mu_{iA}} 
+ \frac{f_2}{y_2+\mu_{jA}}\right] 
+ \frac{ d_i d_j c_i c_j} {2 \mu_Z (y_1+\mu_{iA})}
\left[
\frac{f_1}{y_1+\mu_{iZ}}
+\frac{f_2}{y_2+\mu_{jZ}} \right] 
\non\\
&+& \!\!\!
\Big\{ (y_1,\mu_i) \leftrightarrow (y_2,\mu_j) \Big\} 
\eeq}
\hspace{-0.3cm} with 
{\footnotesize
\beq
{\mathfrak z}_{ij} = 
 \left[ \frac{ d_1 \lambda_{hH_iH_j}}{y_3 - \mu_{1Z}} + 
\frac{ d_2 \lambda_{HH_iH_j}}{y_3 - \mu_{2Z}} \right] + 
\frac{2 d_i d_j}{y_1+\mu_{iZ}} + \frac{2 d_i d_j}{y_2+\mu_{jZ}}
+ \frac{\delta_{ij}}{\mu_Z}
\eeq}
\hspace{-0.3cm} The expressions for $f_{0}$ to $f_3$ have been denoted in
Ref.~\cite{muhl}. The modifications of the SM Higgs-gauge coupling in
the MSSM are accounted for by the mixing parameters:
\beq
\begin{array}{l@{:\;\;}l@{=\;}l l@{:\;\;}l@{=\;}l}
VVh & d_1 & \sin(\beta-\alpha)\;\;\;\; & VVH & d_2 & 
\cos(\beta-\alpha) \\
VAh & c_1 & \cos(\beta-\alpha) \;\;\;\;& VAH & c_2 & 
-\sin(\beta-\alpha)
\end{array}
\eeq
for $V=Z$ and $W$. The reduction of the $Zhh$ cross section is partly
compensated by the $ZHh$ and $ZHH$ cross sections so that their sum
adds up approximately to the SM value, if kinematically possible, as
demonstrated in Fig.~6a for tan $\beta=3$ at $\sqrt{s}=500$~GeV and
$hh$, $Hh$ and $HH$ final states. \pskip

\subsubsection*{3.2 Triple-Higgs Production}

The 2-particle processes \ee $\to ZH_i$ and \ee $\to AH_i$ are among
themselves and mutually complementary to each other in the MSSM
\cite{djoukal}, coming with the coefficients $\sin^2
(\beta-\alpha)/\cos^2(\beta-\alpha)$ and
$\cos^2(\beta-\alpha)/\sin^2(\beta-\alpha)$ for $H_i=h,$ $H$,
respectively. Since multi-Higgs final states are mediated by virtual
$h,$ $H$ bosons, the two types of self-complementarity and mutual
complementarity are also operative in double-Higgs production: \ee
$\to ZH_i H_j,$ $ZAA$ and $AH_i H_j,$ $AAA$. As the different
mechanisms are intertwined, the complementarity between these
3-particle final states is of more complex matrix form, as evident
from Fig.~\ref{fig:graphs}. \s

We will analyze in this section {\it in detail} the processes
involving only the light neutral Higgs boson $h$, \ee $\to Ahh$. The
more cumbersome analyses for heavy neutral Higgs bosons $H$ and other
channels have been presented in Ref.~\cite{muhl}. \s

In the first case one finds for the unpolarized cross section 
\beq
\frac{d\sigma [e^+ e^- \to Ahh]}{dx_1 dx_2} = 
\frac{G_F^3 M_Z^6}{768 \sqrt{2} \pi^3 s} 
\frac{v_e^2 + a_e^2}{(1-\mu_Z)^2} {\mathfrak A}_{11}(x_1,x_2)
\eeq
where the function ${\mathfrak A}_{11}$ reads
\begin{figure}
\begin{center}
\epsfig{figure=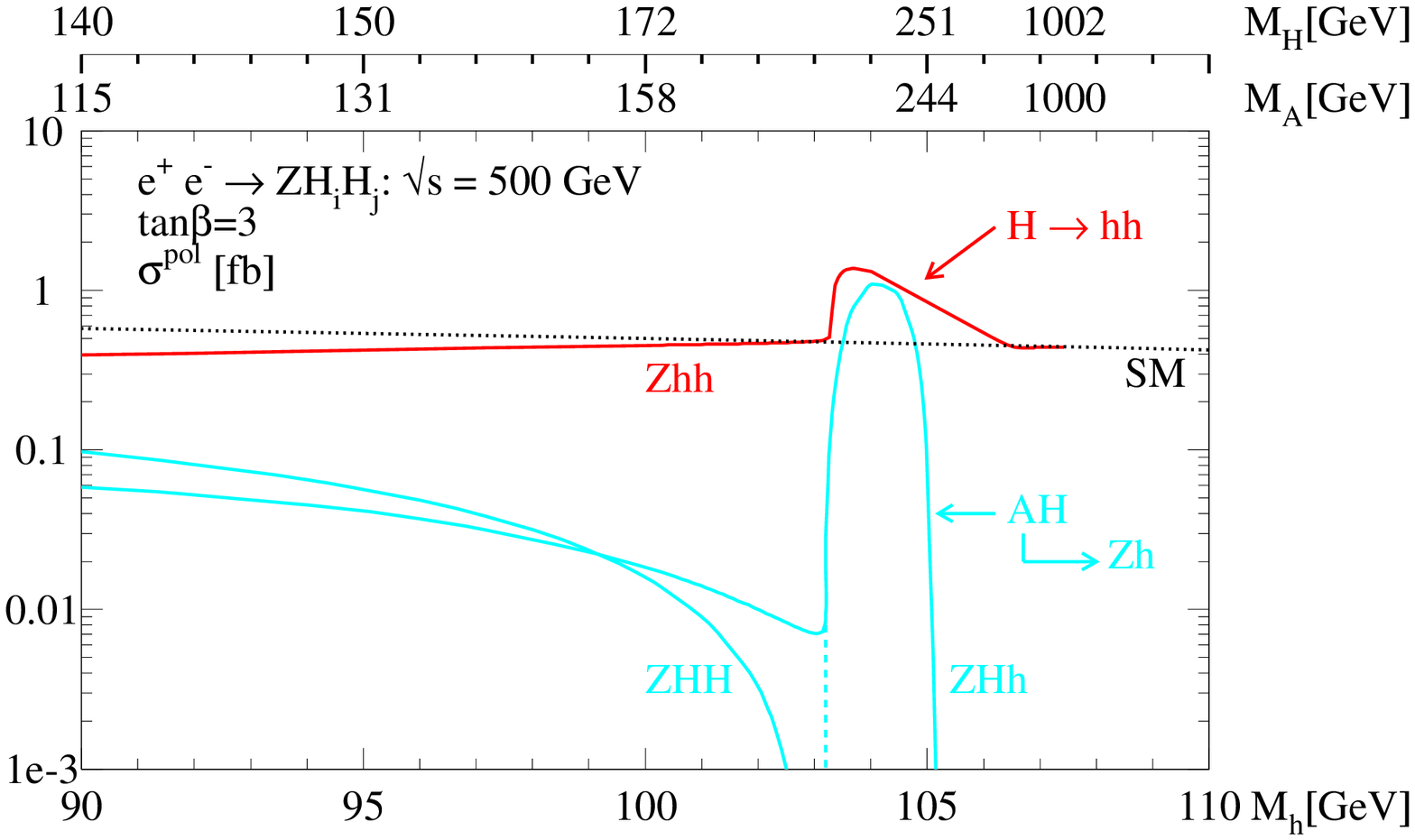,width=13cm}
\\
\end{center}
Figure 6a: {\it Cross sections for the processes $Zhh$, $ZHh$ and 
$ZHH$ for $\sqrt{s}=500$~GeV and tan$\beta=3$, including mixing 
effects ($A = 1$~TeV, $\mu=-1$~TeV).}
\end{figure}
\begin{figure}
\begin{center}
\epsfig{figure=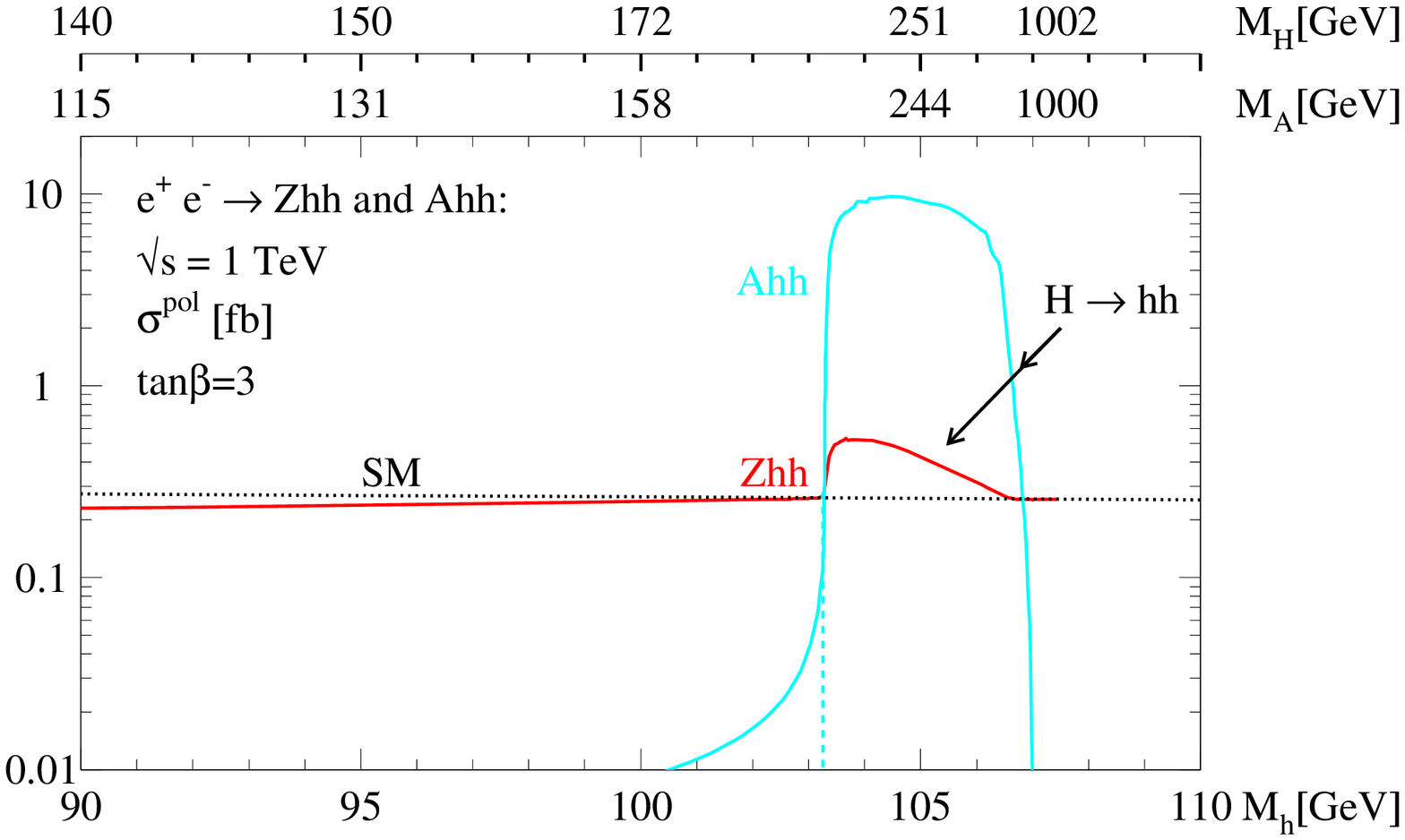,width=13cm}\\[0.9cm]
\end{center}
Figure 6b: {\it Cross sections of the processes Zhh and Ahh for 
$\tan\beta = 3$ and $\sqrt{s}=1$~TeV, including mixing effects 
($A=1$~TeV, $\mu=-1$~TeV).}
\label{fig:ahh}
\end{figure}
\setcounter{figure}{6}
{\footnotesize
\beq
{\mathfrak A}_{11} &=& \left[ 
\frac{c_1 \lambda_{hhh}} {y_3-\mu_{1A}}
+ \frac{c_2 \lambda_{H hh}} {y_3-\mu_{2A}} \right]^2 \frac{g_0}{2}
+ \frac{c_1^2 \lambda^2_{hAA}} {(y_1+\mu_{1A})^2} g_1
+ \frac{c_1^2 d_1^2 } {(y_1+\mu_{1Z})^2} g_2 
\non \\
&+& \left[ \frac{c_1 \lambda_{hhh} } {y_3-\mu_{1A}}
+ \frac{c_2 \lambda_{H hh}} {y_3-\mu_{2A}} \right]
\left[ \frac{c_1 \lambda_{hAA}} {y_1+\mu_{1A}} g_3
+ \frac{c_1 d_1 } {y_1+\mu_{1Z}} g_4 \right] 
+ \frac{c_1^2\lambda_{h AA}^2}{2(y_1+\mu_{1A})(y_2+\mu_{1A})}
g_5 \non\\
&+& \frac{c_1^2 d_1 \lambda_{h AA}}{(y_1+\mu_{1A})(y_1+\mu_{1Z})}g_6
+\frac{c_1^2 d_1 \lambda_{h AA}}{(y_1+\mu_{1A})(y_2+\mu_{1Z})}g_7
+ \frac{c_1^2 d_1^2}{2(y_1+\mu_{1Z})(y_2+\mu_{1Z})}g_8
\non \\
&+& \Big\{ y_1 \leftrightarrow y_2 \Big\}
\eeq}
with $\mu_{1,2} = M_{h,H}^2/s$. The coefficients $g_k$ are listed in 
Ref.~\cite{muhl}.\s

The general form of the double differential cross section of the
process $e^+ e^- \to A H_i H_j$ for unpolarized beams reads in the
same notation as above:
\beq 
\frac{d\sigma [e^+e^- \to AH_iH_j]}{dx_1 dx_2} = 
\frac{G_F^3 M_Z^6}{768 \sqrt{2} \pi^3 s} \, 
\frac{v_e^2+a_e^2}{(1-\mu_Z)^2} \, {\mathfrak A}_{ij}(x_i,x_2) 
\eeq
with the function ${\mathfrak A}_{ij}(x_1,x_2)$ defined by

\vspace{-0.5cm}
{\footnotesize \beq
{\mathfrak A}_{ij} &=&  \left[ 
\frac{ \lambda_{hH_iH_j} c_1} {y_3-\mu_{1A}}
+ \frac{ \lambda_{H H_iH_j} c_2} {y_3-\mu_{2A}} \right]^2 g_0
+ \frac{\lambda^2_{H_jAA} c_i^2} {(y_1+\mu_{iA})^2} g_1
+ \frac{ \lambda^2_{H_iAA} c_j^2} {(y_2+\mu_{jA})^2} g_1' \non \\
&& + \frac{c_j^2 d_i^2 } {(y_1+\mu_{iZ})^2} g_2 +
\frac{c_i^2 d_j^2 } {(y_2+\mu_{jZ})^2} g_2'+
\left[ \frac{\lambda_{hH_iH_j} c_1 } {y_3-\mu_{1A}}
+ \frac{ \lambda_{H H_iH_j} c_2 } {y_3-\mu_{2A}} \right] \non \\
&& \times 
\left[ \frac{ \lambda_{H_jAA} c_i} {y_1+\mu_{iA}} g_3
+ \frac{ \lambda_{H_iAA} c_j} {y_2+\mu_{jA}} g_3' 
+ \frac{c_j d_i } {y_1+\mu_{iZ}} g_4 +
\frac{c_i d_j } {y_2+\mu_{jZ}} g_4' \right] \non \\
&&+ 
\frac{\lambda_{H_i AA} \lambda_{H_j AA} c_ic_j}{(y_1+\mu_{iA})(y_2+\mu_{jA})}g_5
+ \frac{c_ic_j d_i d_j}{(y_1+\mu_{iZ})(y_2+\mu_{jZ})}g_8 
+ \frac{ \lambda_{H_j AA} c_ic_j d_i}{(y_1+\mu_{iA})(y_1+\mu_{iZ})}g_6
\non\\
&& + \frac{\lambda_{H_i AA} c_ic_j d_j}{(y_2+\mu_{jA})(y_2+\mu_{jZ})}
g_6' 
+ \frac{\lambda_{H_j AA} c_i^2 d_j}{(y_1+\mu_{iA})(y_2+\mu_{jZ})} 
g_7 
+ \frac{\lambda_{H_i AA} c_j^2 d_i}{(y_2+\mu_{jA})(y_1+\mu_{iZ})}
g_7'
\eeq}
The coefficients $g_k$ were given in Ref.~\cite{muhl}.\s

The size of the total cross section $\sigma[e^+ e^-\to Ahh]$ is
compared with double Higgs-strahlung $\sigma [e^+ e^-\to Zhh]$ in
Fig.~6b for tan $\beta = 3$ at $\sqrt{s}= 1$~TeV. The cross section
involving the pseudoscalar Higgs boson is small in the continuum. The
effective coupling in the chain $Ah_{virt} \to Ahh$ is
$\cos(\beta-\alpha) \lambda_{hhh}$ while in the chain $AH_{virt} \to
Ahh$ it is $\sin(\beta -\alpha) \lambda_{Hhh}$; both products are
small either in the first or in the second coefficient. Only for
resonance $H$ decays $AH \to Ahh$ the cross section becomes very
large as expected from the decoupling theorem. \pskip

\subsubsection*{3.4 Sensitivity Areas}

The results obtained in the preceding sections can be summarized in
compact form by constructing sensitivity areas for the trilinear SUSY
Higgs couplings based on the cross sections for double Higgs-strahlung
and triple Higgs production. $WW$ double-Higgs fusion can provide
additional information on the Higgs self-couplings, in particluar for
large collider energies. \s

The sensitivity areas will be defined in the $[M_A,$ tan$\beta]$ plane
\cite{djouadi}. The criteria for accepting a point in the plane as
accessible for the measurement of a specific trilinear coupling
are set as follows:
\beq
\begin{array}{l l} 
(i) & \sigma [\lambda] > 0.01~{\rm fb} \\
(ii) & {\rm eff}\{ \lambda \to 0 \} > 
2~{\rm st.dev.} \quad {\rm for} \quad \int {\cal L} = 2~{\rm ab}^{-1}
\end{array} 
\label{crit}
\eeq
The first criterion demands at least 20 events in a sample collected
for an integrated lu\-mi\-no\-si\-ty of 2~ab$^{-1}$, corresponding to 
the lifetime of a high-luminosity machine such as TESLA. The
second criterion demands at least a 2 standard-deviation effect of the
non-zero trilinear coupling away from zero. Even though the two
criteria may look quite loose, the tightening of $(i)$ and/or $(ii)$
does not have a large impact on the size of the sensitivity areas in
the $[M_A,$ tan$\beta]$ plane, see Ref.~\cite{osland}. The second
criterion was defined in Ref.~\cite{muhl} slightly different by
introducing the relative variation of the trilinear couplings with
respect to the MSSM. [Due to an algorithmic error, the sensitivity
areas in Ref.~\cite{muhl} had been overestimated.]  For the sake of
simplicity, mixing effects are neglected in the analysis.\s

Sensitivity areas of the trilinear couplings among the scalar Higgs
bosons $h$, $H$ in the correlation matrix Table~\ref{tab:coup}, are
depicted in Figs.~7a and 7b.  If at most one heavy
Higgs boson is present in the final state, the lower energy
$\sqrt{s}=500$~GeV is more preferable in the case of double
Higgs-strahlung. $HH$ final states in double Higgs-strahlung and
triple Higgs production including $A$ give rise to larger sensitivity
areas at the high energy $\sqrt{s}=1$~TeV. Apart from small regions in
which interference effects play a major role, the magnitude of the
sensitivity regions in the parameter tan$\beta$ is readily explained
by the magnitude of the parameters $\lambda \sin (\beta-\alpha)$ and
$\lambda \cos (\beta-\alpha)$, shown individually in
Figs.~2a and 2b. For large $M_A$ the
sensitivity criteria cannot be met any more either as a result of
phase space effects or due to the suppression of the $H$, $A$ 
propagators for large masses. While the trilinear coupling of the
light neutral CP-even Higgs boson is accessible in nearly the entire
MSSM parameter space, the regions for the $\lambda$'s involving heavy
Higgs bosons are rather restricted.\s

\begin{figure}
\begin{center}
\epsfig{figure=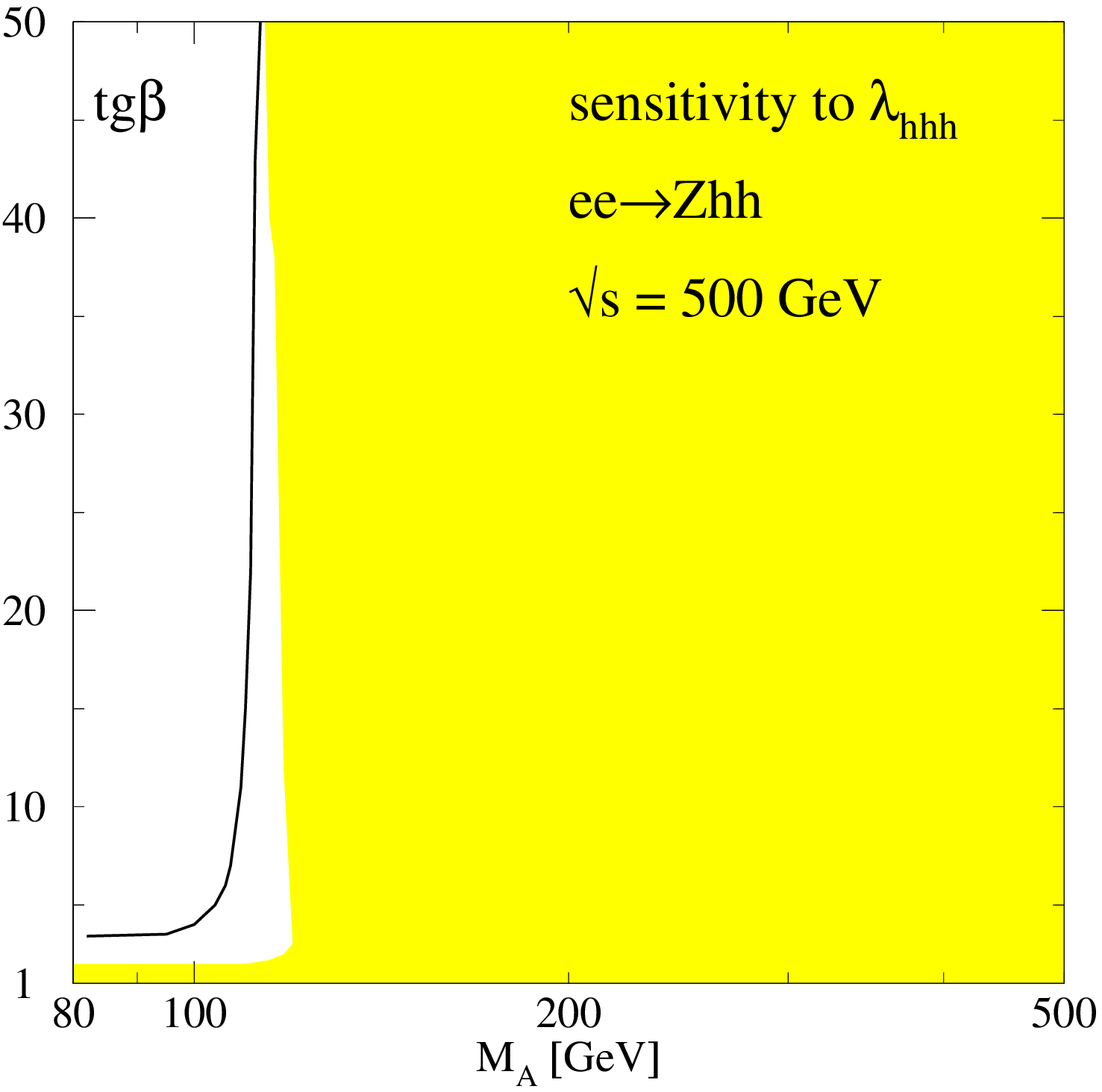,width=7cm}
\hspace{1cm}
\epsfig{figure=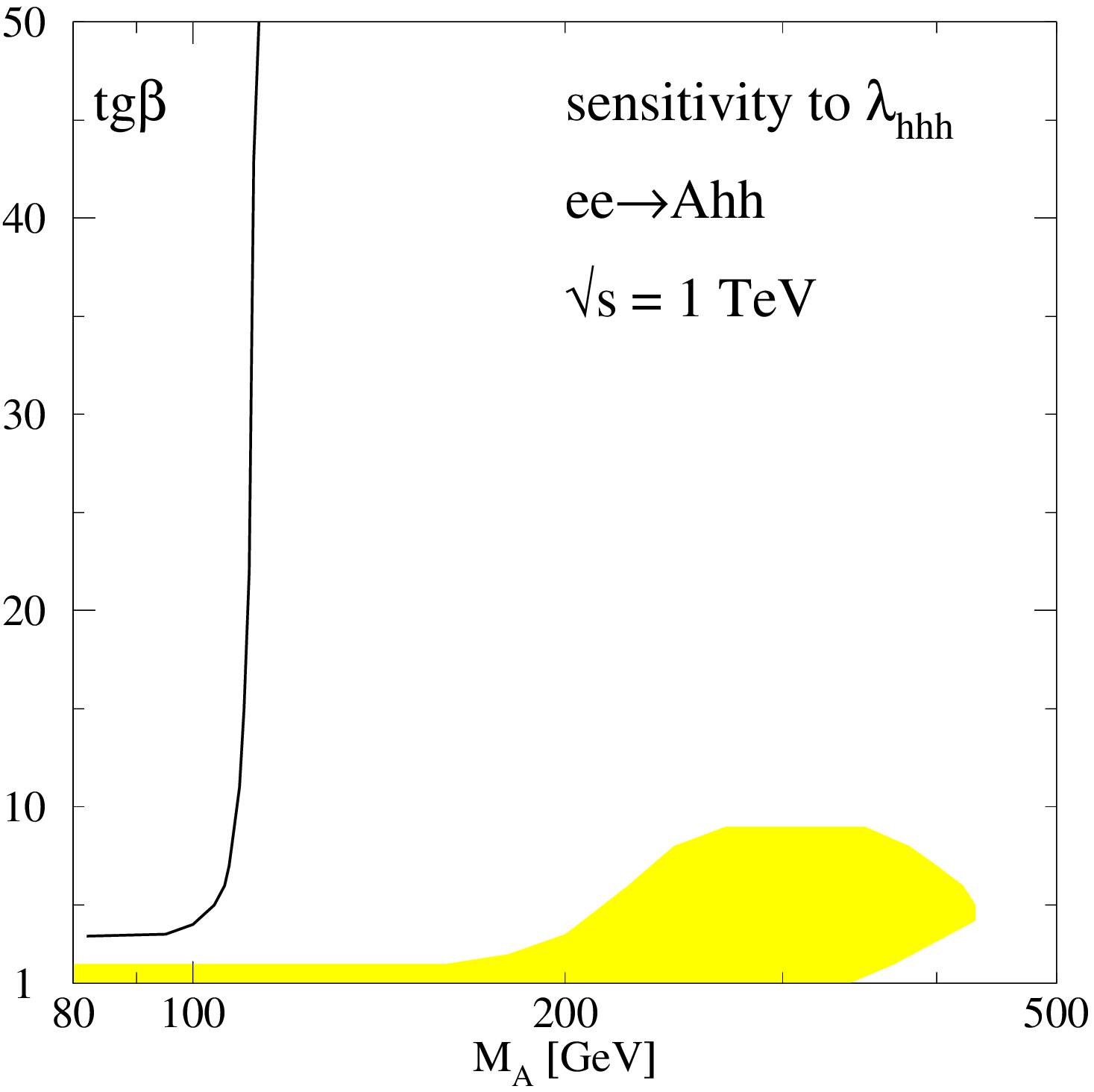,width=7cm}
\end{center}
\vspace{1.5cm}
\begin{center}
\epsfig{figure=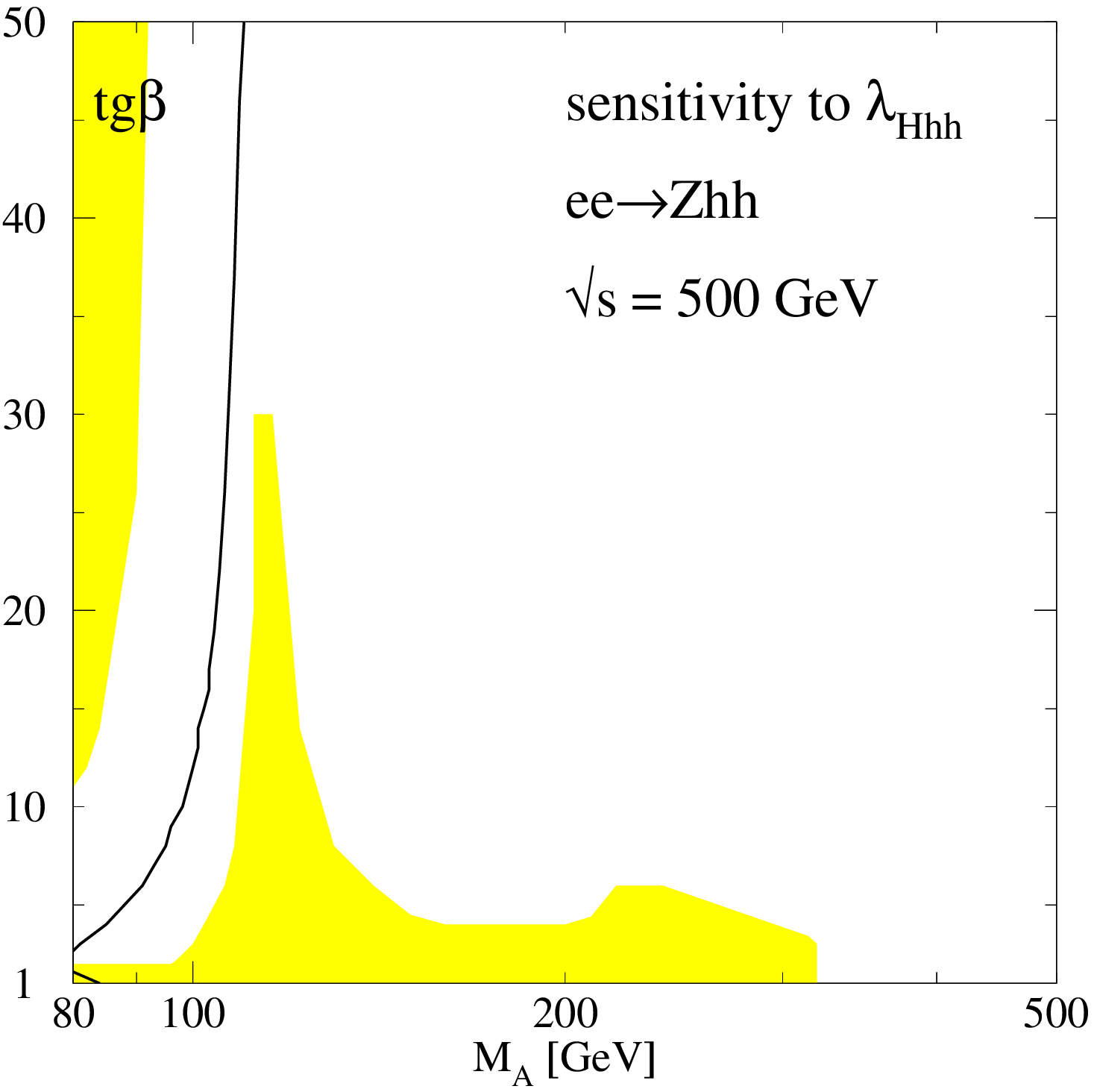,width=7cm}
\hspace{1cm}
\epsfig{figure=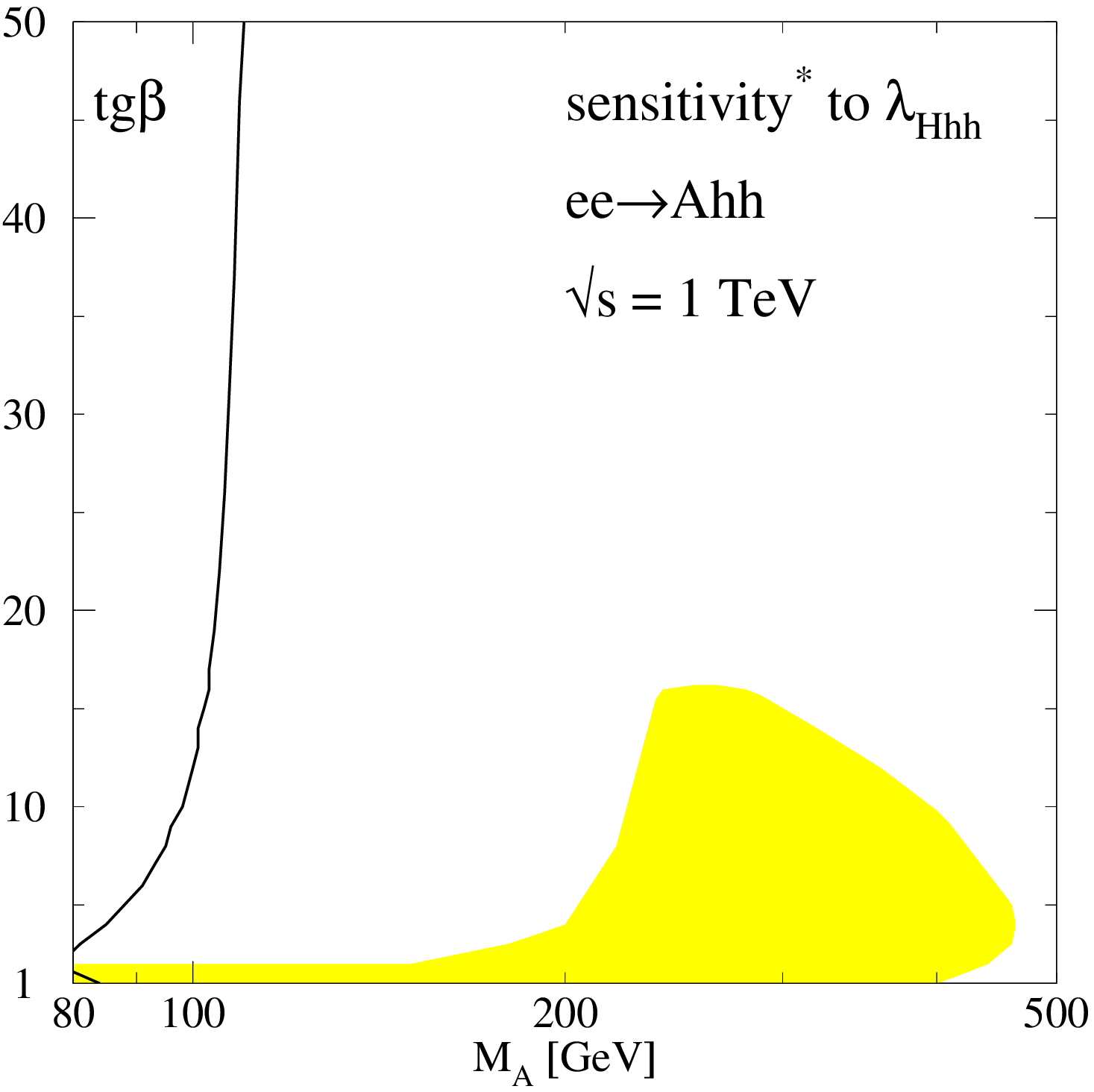,width=7cm}\\[0.5cm]
\label{fig:s1}
\end{center}
Figure 7a: {\it Sensitivity [* def = Ref.\cite{muhl}] to the couplings 
$\lambda_{hhh}$ and $\lambda_{Hhh}$ in the 
processes \ee$\to Zhh$ and \ee$\to Ahh$ for collider energies 
$500$~GeV and $1$~TeV, respectively (no mixing). [Vanishing trilinear 
couplings are indicated by contour lines.]}
\end{figure}
\begin{figure}
\begin{center}
\epsfig{figure=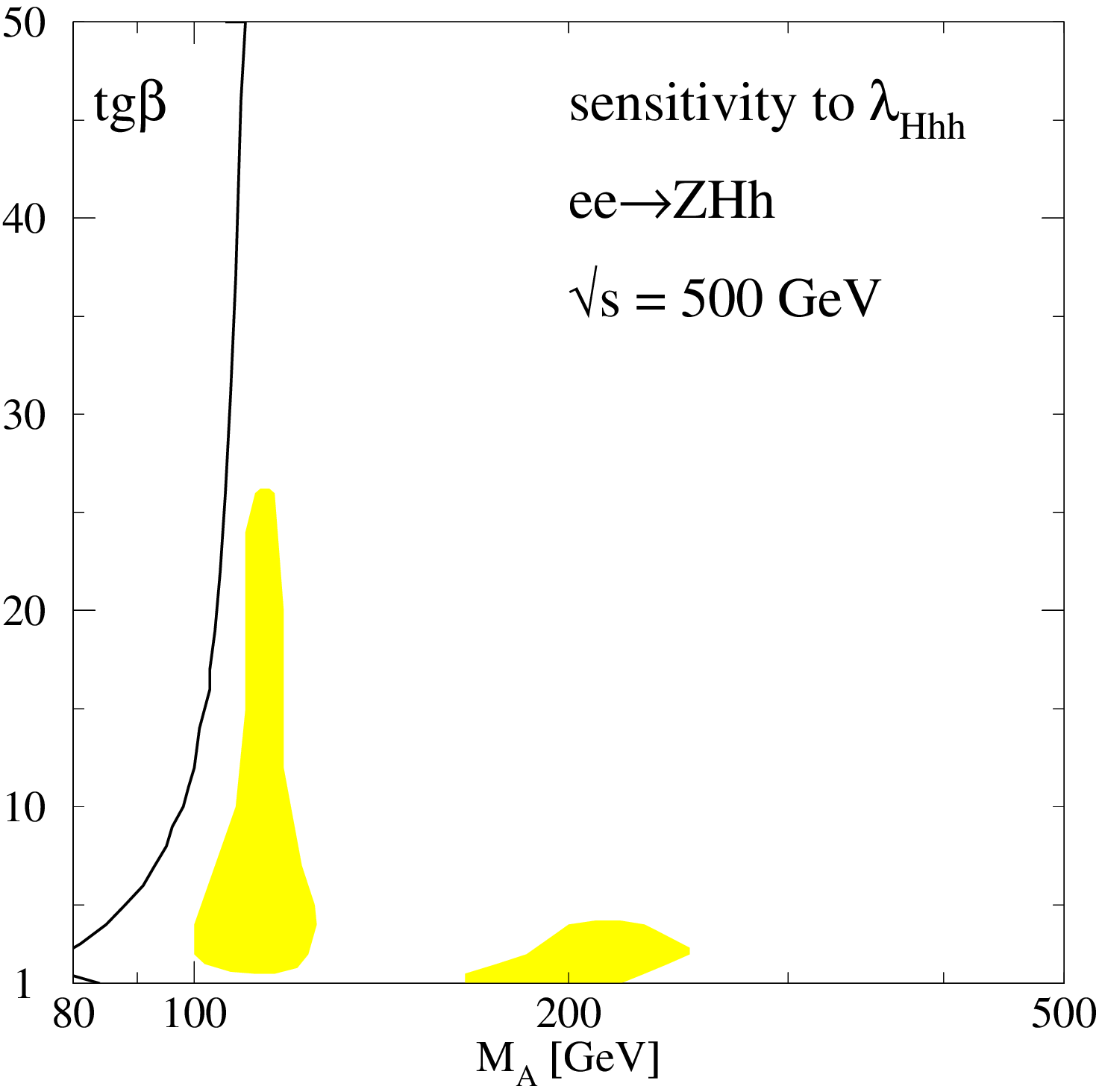,width=7cm}
\hspace{1cm}
\epsfig{figure=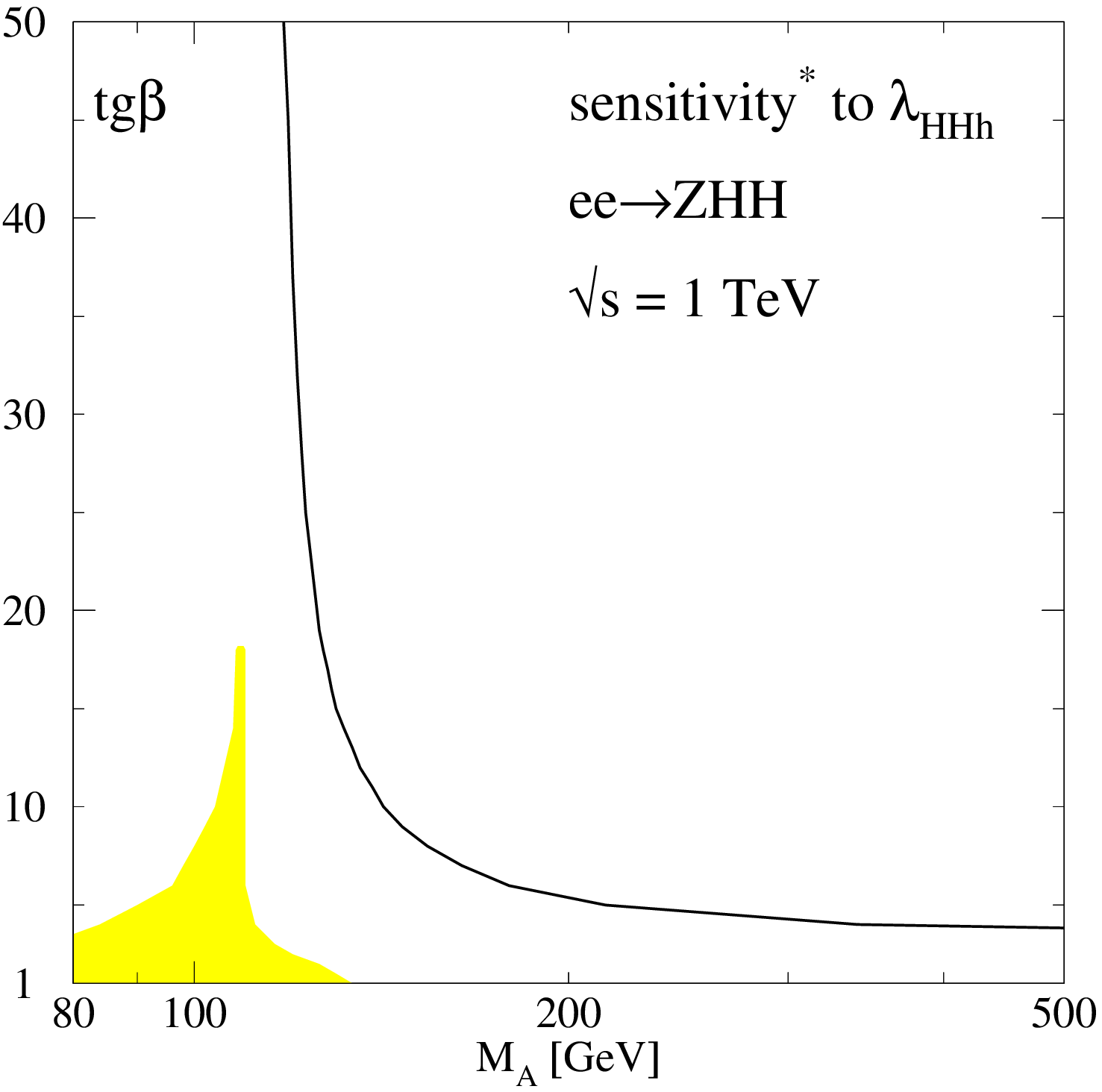,width=7cm}
\end{center}
\vspace{1.5cm}
\begin{center}
\epsfig{figure=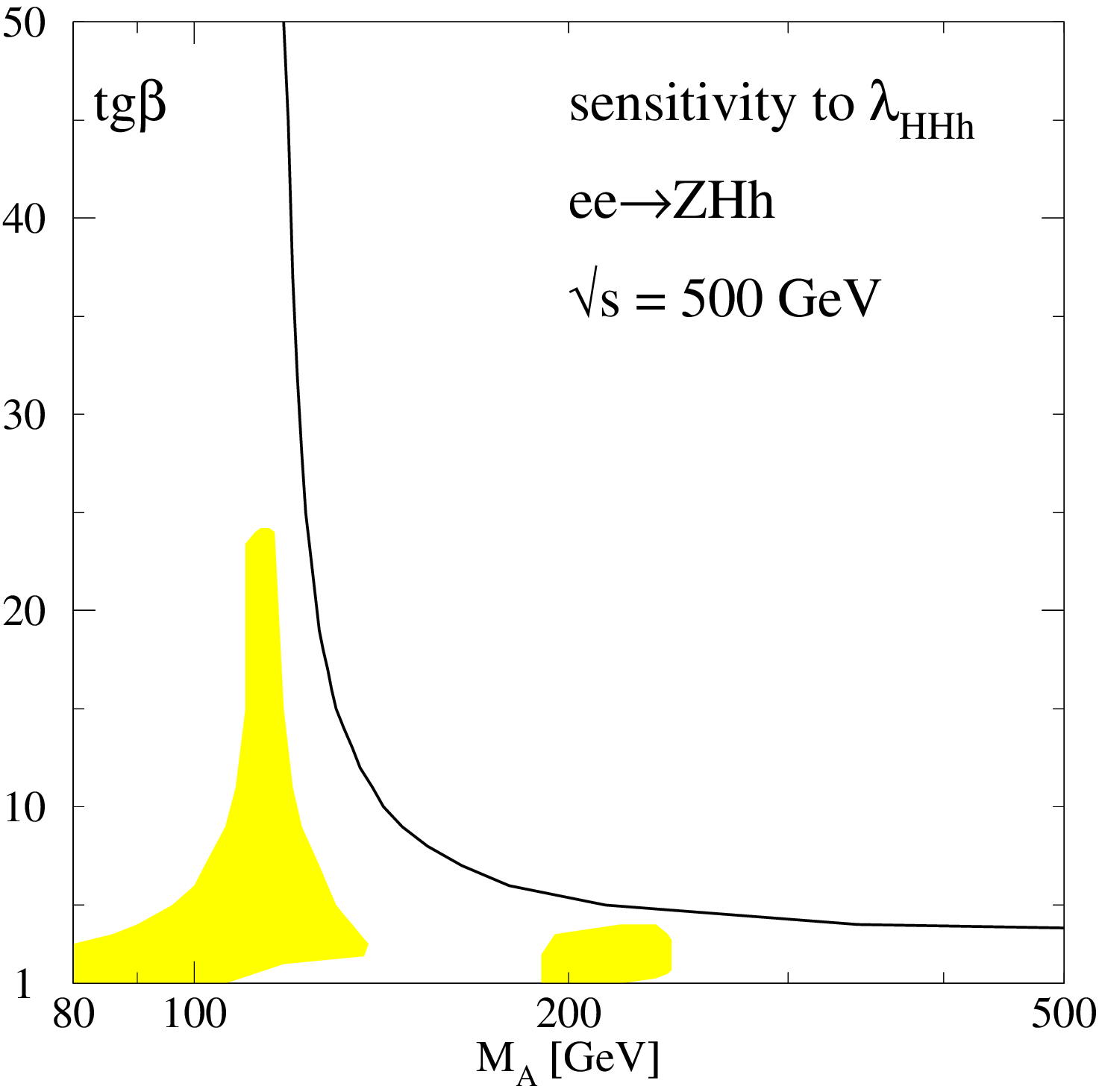,width=7cm}
\hspace{1cm}
\epsfig{figure=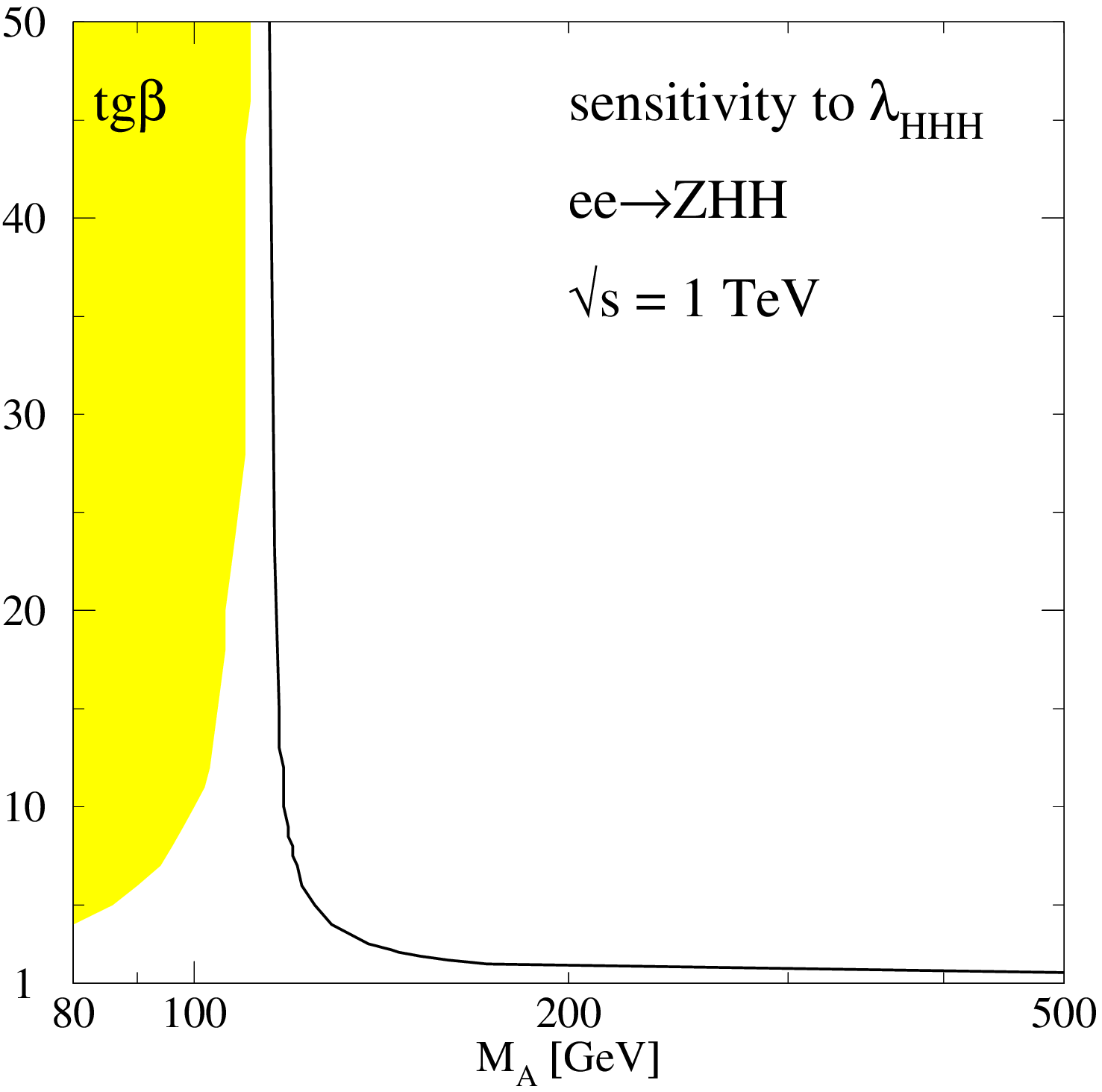,width=7cm}\\[0.5cm]
\label{fig:s2}
\end{center}
Figure 7b: {\it Sensitivity [* eff$\{\lambda\to 0\}>1$st.dev.] to 
the couplings $\lambda_{Hhh}$, $\lambda_{HHh}$ and $\lambda_{HHH}$ 
in the processes \ee$\to ZHh$ and \ee$\to ZHH$ for collider energies 
$500$~GeV and $1$~TeV, respectively (no mixing).}
\end{figure}
Since neither experimental efficiencies nor background related cuts
are considered in this paper, the areas shown in Figs.~7a
and 7b must be interpreted as maximal envelopes. They are
expected to shrink when experimental efficiencies are properly taken
into account; more elaborate cuts on signal and backgrounds, however,
may help reduce their impact.\pskip

%%%%%%%%%%%%%%%%%%%%%%%%%%%%%%%%%%%%%%%%%%%%%%%%%%%%%%%%%%%%%%%%%%%%%%%%
\subsection*{4. Conclusions}

In this report we have analyzed the production of Higgs boson pairs
and triple Higgs final states at $e^+ e^-$ linear colliders up to
energies of 1~TeV.  They will allow the measurement of the fundamental
trilinear Higgs self-couplings. The first theoretical steps into this
area have been taken by calculating the production cross sections in
the Standard Model for Higgs bosons in the intermediate mass range and
for Higgs bosons in the minimal supersymmetric extension. \s

The cross sections in the Standard Model for double Higgs-strahlung
and triple Higgs pro\-duc\-tion are small so that high luminosities
are needed to perform these experiments. Even though the $e^+ e^-$
cross sections are smaller than the corresponding $pp$ cross sections,
the strongly reduced number of background events renders the search
for the Higgs-pair signal events, through $bbbb$ final states for
instance, easier in the $e^+ e^-$ environment than in jetty LHC final
states.  For sufficiently high luminosities even the first phase of
these colliders with an energy of 500 GeV will allow the experimental
analysis of self-couplings for Higgs bosons in the intermediate mass
range. \s

The extended Higgs spectrum in supersymmetric theories gives rise to a
plethora of trilinear and quadrilinear couplings. The $hhh$ coupling
is generally quite different from the Standard Model. It can be
measured in $hh$ continuum production at $e^+e^-$ linear colliders.
Other couplings between heavy and light scalar Higgs bosons can be
measured as well, though only in restricted areas of the $[M_A,$
tan$\beta]$ parameter space as illustrated in the set of Figs.~7a and
7b. The trilinear couplings including the pseudoscalar Higgs boson $A$
are predicted to be small in the MSSM; future experimental analyses
are therefore expected to give rise to upper bounds on these couplings
if the MSSM is realized in Nature. For more general supersymmetric
theories a model-independent analysis must be performed in triple
Higgs channels.\pskip

\subsubsection*{Acknowledgements}

We gratefully acknowledge discussions with P.~Gay, P.~Lutz, L.~Maiani,
P.~Osland and F.~Richard.

\vspace{4cm}

%%%%%%%%%%%%%%%%%%%%%%%%%%%%%%%%%%%%%%%%%%%%%%%%%%%%%%%%%%%%%%%%%%%%%%%%
%%% References
%%%%%%%%%%%%%%%%%%%%%%%%%%%%%%%%%%%%%%%%%%%%%%%%%%%%%%%%%%%%%%%%%%%%%%%%

\end{document}